%
%
%
%
\def\unredoffs{} \def\redoffs{\voffset=-.40truein\hoffset=-.40truein}
\def\speclscape{}
%
%
%
%
\newbox\leftpage \newdimen\fullhsize \newdimen\hstitle \newdimen\hsbody
\tolerance=1000\hfuzz=2pt
\catcode`\@=11 
\def\bigans{b }
\def\answ{b }
\ifx\answ\bigans\message{(This will come out unreduced.}
\magnification=1200\unredoffs\baselineskip=16pt plus 2pt minus 1pt
\hsbody=\hsize \hstitle=\hsize 
\else\message{(This will be reduced.} \let\l@r=L
\magnification=1000\baselineskip=16pt plus 2pt minus 1pt \vsize=7truein
\redoffs \hstitle=8truein\hsbody=4.75truein\fullhsize=10truein\hsize=\hsbody
\output={\ifnum\pageno=0 
    \shipout\vbox{\speclscape{\hsize\fullhsize\makeheadline}
      \hbox to \fullhsize{\hfill\pagebody\hfill}}\advancepageno
    \else
    \almostshipout{\leftline{\vbox{\pagebody\makefootline}}}\advancepageno
    \fi}
\def\almostshipout#1{\if L\l@r \count1=1 \message{[\the\count0.\the\count1]}
        \global\setbox\leftpage=#1 \global\let\l@r=R
   \else \count1=2
    \shipout\vbox{\speclscape{\hsize\fullhsize\makeheadline}
        \hbox to\fullhsize{\box\leftpage\hfil#1}}  \global\let\l@r=L\fi}
\fi
%
\newcount\yearltd\yearltd=\year\advance\yearltd by -1900

\def\Title#1#2{\nopagenumbers\abstractfont\hsize=\hstitle\rightline{#1}%
\vskip 1in\centerline{\titlefont #2}\abstractfont\vskip .5in\pageno=0}
%
%

\def\draftmode{\message{ DRAFTMODE }\def\draftdate{{\rm preliminary draft:
\number\month/\number\day/\number\yearltd\ \ \hourmin}}%
\headline={\hfil\draftdate}\writelabels\baselineskip=20pt plus 2pt minus 2pt
   {\count255=\time\divide\count255 by 60 \xdef\hourmin{\number\count255}
    \multiply\count255 by-60\advance\count255 by\time
    \xdef\hourmin{\hourmin:\ifnum\count255<10 0\fi\the\count255}}}
\def\nolabels{\def\wrlabeL##1{}\def\eqlabeL##1{}\def\reflabeL##1{}}
\def\writelabels{\def\wrlabeL##1{\leavevmode\vadjust{\rlap{\smash%
{\line{{\escapechar=` \hfill\rlap{\sevenrm\hskip.03in\string##1}}}}}}}%
\def\eqlabeL##1{{\escapechar-1\rlap{\sevenrm\hskip.05in\string##1}}}%
\def\reflabeL##1{\noexpand\llap{\noexpand\sevenrm\string\string\string##1}}}
\nolabels
%
\global\newcount\secno \global\secno=0
\global\newcount\meqno \global\meqno=1
\def\newsec#1{\global\advance\secno by1\message{(\the\secno. #1)}
\global\subsecno=0\eqnres@t\noindent{\bf\the\secno. #1}
\writetoca{{\secsym} {#1}}\par\nobreak\medskip\nobreak}
\def\eqnres@t{\xdef\secsym{\the\secno.}\global\meqno=1\bigbreak\bigskip}
\def\sequentialequations{\def\eqnres@t{\bigbreak}}\xdef\secsym{}
\global\newcount\subsecno \global\subsecno=0
\def\subsec#1{\global\advance\subsecno by1\message{(\secsym\the\subsecno. #1)}
\ifnum\lastpenalty>9000\else\bigbreak\fi
\noindent{\it\secsym\the\subsecno. #1}\writetoca{\string\quad
{\secsym\the\subsecno.} {#1}}\par\nobreak\medskip\nobreak}
\def\appendix#1#2{\global\meqno=1\global\subsecno=0\xdef\secsym{\hbox{#1.}}
\bigbreak\bigskip\noindent{\bf Appendix #1. #2}\message{(#1. #2)}
\writetoca{Appendix {#1.} {#2}}\par\nobreak\medskip\nobreak}
%
%
\def\eqnn#1{\xdef #1{(\secsym\the\meqno)}\writedef{#1\leftbracket#1}%
\global\advance\meqno by1\wrlabeL#1}
\def\eqna#1{\xdef #1##1{\hbox{$(\secsym\the\meqno##1)$}}
\writedef{#1\numbersign1\leftbracket#1{\numbersign1}}%
\global\advance\meqno by1\wrlabeL{#1$\{\}$}}
\def\eqn#1#2{\xdef #1{(\secsym\the\meqno)}\writedef{#1\leftbracket#1}%
\global\advance\meqno by1$$#2\eqno#1\eqlabeL#1$$}
%
\newskip\footskip\footskip14pt plus 1pt minus 1pt 
\def\footnotefont{\ninepoint}\def\f@t#1{\footnotefont #1\@foot}
\def\f@@t{\baselineskip\footskip\bgroup\footnotefont\aftergroup\@foot\let\next}
\setbox\strutbox=\hbox{\vrule height9.5pt depth4.5pt width0pt}
\global\newcount\ftno \global\ftno=0
\def\foot{\global\advance\ftno by1\footnote{$^{\the\ftno}$}}
%
\newwrite\ftfile
\def\footend{\def\foot{\global\advance\ftno by1\chardef\wfile=\ftfile
$^{\the\ftno}$\ifnum\ftno=1\immediate\openout\ftfile=foots.tmp\fi%
\immediate\write\ftfile{\noexpand\smallskip%
\noexpand\item{f\the\ftno:\ }\pctsign}\findarg}%
\def\footatend{\vfill\eject\immediate\closeout\ftfile{\parindent=20pt
\centerline{\bf Footnotes}\nobreak\bigskip\input foots.tmp }}}
\def\footatend{}
%
%
\global\newcount\refno \global\refno=1
\newwrite\rfile
\def\ref{[\the\refno]\nref}
\def\nref#1{\xdef#1{[\the\refno]}\writedef{#1\leftbracket#1}%
\ifnum\refno=1\immediate\openout\rfile=refs.tmp\fi
\global\advance\refno by1\chardef\wfile=\rfile\immediate
\write\rfile{\noexpand\item{#1\ }\reflabeL{#1\hskip.31in}\pctsign}\findarg}
\def\findarg#1#{\begingroup\obeylines\newlinechar=`\^^M\pass@rg}
{\obeylines\gdef\pass@rg#1{\writ@line\relax #1^^M\hbox{}^^M}%
\gdef\writ@line#1^^M{\expandafter\toks0\expandafter{\striprel@x #1}%
\edef\next{\the\toks0}\ifx\next\em@rk\let\next=\endgroup\else\ifx\next\empty%
\else\immediate\write\wfile{\the\toks0}\fi\let\next=\writ@line\fi\next\relax}}
\def\striprel@x#1{} \def\em@rk{\hbox{}}
\def\lref{\begingroup\obeylines\lr@f}
\def\lr@f#1#2{\gdef#1{\ref#1{#2}}\endgroup\unskip}
\def\semi{;\hfil\break}
\def\addref#1{\immediate\write\rfile{\noexpand\item{}#1}} 
\def\startrefs#1{\immediate\openout\rfile=refs.tmp\refno=#1}
\def\xref{\expandafter\xr@f}\def\xr@f[#1]{#1}
\def\refs#1{\count255=1[\r@fs #1{\hbox{}}]}
\def\r@fs#1{\ifx\und@fined#1\message{reflabel \string#1 is undefined.}%
\nref#1{need to supply reference \string#1.}\fi%
\vphantom{\hphantom{#1}}\edef\next{#1}\ifx\next\em@rk\def\next{}%
\else\ifx\next#1\ifodd\count255\relax\xref#1\count255=0\fi%
\else#1\count255=1\fi\let\next=\r@fs\fi\next}
%

%
\newwrite\ffile\global\newcount\figno \global\figno=1
\def\fig{fig.~\the\figno\nfig}
\def\nfig#1{\xdef#1{fig.~\the\figno}%
\writedef{#1\leftbracket fig.\noexpand~\the\figno}%
\ifnum\figno=1\immediate\openout\ffile=figs.tmp\fi\chardef\wfile=\ffile%
\immediate\write\ffile{\noexpand\medskip\noexpand\item{Fig.\ \the\figno. }
\reflabeL{#1\hskip.55in}\pctsign}\global\advance\figno by1\findarg}
\def\xfig{\expandafter\xf@g}\def\xf@g fig.\penalty\@M\ {}
\def\figs#1{figs.~\f@gs #1{\hbox{}}}
\def\f@gs#1{\edef\next{#1}\ifx\next\em@rk\def\next{}\else
\ifx\next#1\xfig #1\else#1\fi\let\next=\f@gs\fi\next}
\newwrite\lfile
{\escapechar-1\xdef\pctsign{\string\%}\xdef\leftbracket{\string\{}
\xdef\rightbracket{\string\}}\xdef\numbersign{\string\#}}

\def\writestop{\def\writestoppt{\immediate\write\lfile{\string\pageno%
\the\pageno\string\startrefs\leftbracket\the\refno\rightbracket%
\string\def\string\secsym\leftbracket\secsym\rightbracket%
\string\secno\the\secno\string\meqno\the\meqno}\immediate\closeout\lfile}}
\def\writestoppt{}\def\writedef#1{}
\def\seclab#1{\xdef #1{\the\secno}\writedef{#1\leftbracket#1}\wrlabeL{#1=#1}}
\def\subseclab#1{\xdef #1{\secsym\the\subsecno}%
\writedef{#1\leftbracket#1}\wrlabeL{#1=#1}}
\newwrite\tfile \def\writetoca#1{}
\def\leaderfill{\leaders\hbox to 1em{\hss.\hss}\hfill}
\def\writetoc{\immediate\openout\tfile=toc.tmp
     \def\writetoca##1{{\edef\next{\write\tfile{\noindent ##1
     \string\leaderfill {\noexpand\number\pageno} \par}}\next}}}

%
\catcode`\@=12 
%
\edef\tfontsize{\ifx\answ\bigans scaled\magstep3\else scaled\magstep4\fi}
\font\titlerm=cmr10 \tfontsize \font\titlerms=cmr7 \tfontsize
\font\titlermss=cmr5 \tfontsize \font\titlei=cmmi10 \tfontsize
\font\titleis=cmmi7 \tfontsize \font\titleiss=cmmi5 \tfontsize
\font\titlesy=cmsy10 \tfontsize \font\titlesys=cmsy7 \tfontsize
\font\titlesyss=cmsy5 \tfontsize \font\titleit=cmti10 \tfontsize
\skewchar\titlei='177 \skewchar\titleis='177 \skewchar\titleiss='177
\skewchar\titlesy='60 \skewchar\titlesys='60 \skewchar\titlesyss='60
\def\titlefont{\def\rm{\fam0\titlerm}
\textfont0=\titlerm \scriptfont0=\titlerms \scriptscriptfont0=\titlermss
\textfont1=\titlei \scriptfont1=\titleis \scriptscriptfont1=\titleiss
\textfont2=\titlesy \scriptfont2=\titlesys \scriptscriptfont2=\titlesyss
\textfont\itfam=\titleit \def\it{\fam\itfam\titleit}\rm}
 \ifx\answ\bigans\else scaled\magstep1\fi
\ifx\answ\bigans\def\abstractfont{\tenpoint}\else
\font\abssl=cmsl10 scaled \magstep1
\font\absrm=cmr10 scaled\magstep1 \font\absrms=cmr7 scaled\magstep1
\font\absrmss=cmr5 scaled\magstep1 \font\absi=cmmi10 scaled\magstep1
\font\absis=cmmi7 scaled\magstep1 \font\absiss=cmmi5 scaled\magstep1
\font\abssy=cmsy10 scaled\magstep1 \font\abssys=cmsy7 scaled\magstep1
\font\abssyss=cmsy5 scaled\magstep1 \font\absbf=cmbx10 scaled\magstep1
\skewchar\absi='177 \skewchar\absis='177 \skewchar\absiss='177
\skewchar\abssy='60 \skewchar\abssys='60 \skewchar\abssyss='60
\def\abstractfont{\def\rm{\fam0\absrm}
\textfont0=\absrm \scriptfont0=\absrms \scriptscriptfont0=\absrmss
\textfont1=\absi \scriptfont1=\absis \scriptscriptfont1=\absiss
\textfont2=\abssy \scriptfont2=\abssys \scriptscriptfont2=\abssyss
\textfont\itfam=\bigit \def\it{\fam\itfam\bigit}\def\footnotefont{\tenpoint}%
\textfont\slfam=\abssl \def\sl{\fam\slfam\abssl}%
\textfont\bffam=\absbf \def\bf{\fam\bffam\absbf}\rm}\fi
\def\tenpoint{\def\rm{\fam0\tenrm}
\textfont0=\tenrm \scriptfont0=\sevenrm \scriptscriptfont0=\fiverm
\textfont1=\teni  \scriptfont1=\seveni  \scriptscriptfont1=\fivei
\textfont2=\tensy \scriptfont2=\sevensy \scriptscriptfont2=\fivesy
\textfont\itfam=\tenit \def\it{\fam\itfam\tenit}\def\footnotefont{\ninepoint}%
\textfont\bffam=\tenbf \def\bf{\fam\bffam\tenbf}\def\sl{\fam\slfam\tensl}\rm}
\font\ninerm=cmr9 \font\sixrm=cmr6 \font\ninei=cmmi9 \font\sixi=cmmi6
\font\ninesy=cmsy9 \font\sixsy=cmsy6 \font\ninebf=cmbx9
\font\nineit=cmti9 \font\ninesl=cmsl9 \skewchar\ninei='177
\skewchar\sixi='177 \skewchar\ninesy='60 \skewchar\sixsy='60
\def\ninepoint{\def\rm{\fam0\ninerm}
\textfont0=\ninerm \scriptfont0=\sixrm \scriptscriptfont0=\fiverm
\textfont1=\ninei \scriptfont1=\sixi \scriptscriptfont1=\fivei
\textfont2=\ninesy \scriptfont2=\sixsy \scriptscriptfont2=\fivesy
\textfont\itfam=\ninei \def\it{\fam\itfam\nineit}\def\sl{\fam\slfam\ninesl}%
\textfont\bffam=\ninebf \def\bf{\fam\bffam\ninebf}\rm}
%
%

\hyphenation{anom-aly anom-alies coun-ter-term coun-ter-terms}
\def\inv{^{\raise.15ex\hbox{${\scriptscriptstyle -}$}\kern-.05em 1}}

\def\Dsl{\,\raise.15ex\hbox{/}\mkern-13.5mu D} 
\def\dsl{\raise.15ex\hbox{/}\kern-.57em\partial}

 \def\Tr{{\rm Tr}}
\font\bigit=cmti10 scaled \magstep1
\def\lspace{\ifx\answ\bigans{}\else\qquad\fi}
\def\lbspace{\ifx\answ\bigans{}\else\hskip-.2in\fi} 
\def\boxeqn#1{\vcenter{\vbox{\hrule\hbox{\vrule\kern3pt\vbox{\kern3pt
      \hbox{${\displaystyle #1}$}\kern3pt}\kern3pt\vrule}\hrule}}}
\def\mbox#1#2{\vcenter{\hrule \hbox{\vrule height#2in
          \kern#1in \vrule} \hrule}}  
%
 \def\CO{{\cal O}} 
\def\CA{{\cal A}}  \def\CF{{\cal F}} 
\def\CL{{\cal L}} \def\CH{{\cal H}}  \def\CU{{\cal U}}
  \def\CD{{\cal D}} 
\def\e#1{{\rm e}^{^{\textstyle#1}}}

\def\darr#1{\raise1.5ex\hbox{$\leftrightarrow$}\mkern-16.5mu #1}

\def\half{{\textstyle{1\over2}}} 
\def\roughly#1{\raise.3ex\hbox{$#1$\kern-.75em\lower1ex\hbox{$\sim$}}}

\def\np#1#2#3{Nucl. Phys. {\bf B#1} (#2) #3}
\def\pl#1#2#3{Phys. Lett. {\bf #1B} (#2) #3}

\def\anp#1#2#3{Ann. Phys. {\bf #1} (#2) #3}
\def\pr#1#2#3{Phys. Rev. {\bf #1} (#2) #3}
\def\ap#1#2#3{Ann. Phys. {\bf #1} (#2) #3}

\def\cmp#1#2#3{Comm. Math. Phys. {\bf #1} (#2) #3}
\def\mpl#1#2#3{Mod. Phys. Lett. {\bf #1} (#2) #3}

\def\jhep#1#2#3{JHEP {\bf#1}(#2) #3}

\def\ijmp#1#2#3{Int.~J.~Mod.~Phys. {\bf #1} (#2) #3}
\def\atmp#1#2#3{Adv.~Theor.~Math.~Phys.{\bf #1} (#2) #3}
\def\ap#1#2#3{Ann.~Phys. {\bf #1} (#2) #3}
\def\IB{\relax\hbox{$\inbar\kern-.3em{\rm B}$}}
\def\IC{\relax\hbox{$\inbar\kern-.3em{\rm C}$}}
\def\ID{\relax\hbox{$\inbar\kern-.3em{\rm D}$}}
\def\IE{\relax\hbox{$\inbar\kern-.3em{\rm E}$}}
\def\IF{\relax\hbox{$\inbar\kern-.3em{\rm F}$}}
\def\IG{\relax\hbox{$\inbar\kern-.3em{\rm G}$}}
\def\IGa{\relax\hbox{${\rm I}\kern-.18em\Gamma$}}
\def\IH{\relax{\rm I\kern-.18em H}}
\def\IK{\relax{\rm I\kern-.18em K}}
\def\IL{\relax{\rm I\kern-.18em L}}
\def\IP{\relax{\rm I\kern-.18em P}}
\def\IR{\relax{\rm I\kern-.18em R}}
\def\IZ{\relax\ifmmode\mathchoice{
\hbox{\cmss Z\kern-.4em Z}}{\hbox{\cmss Z\kern-.4em Z}}
{\lower.9pt\hbox{\cmsss Z\kern-.4em Z}}
{\lower1.2pt\hbox{\cmsss Z\kern-.4em Z}}
\else{\cmss Z\kern-.4em Z}\fi}
\def\II{\relax{\rm I\kern-.18em I}}

\def\ndt{{\noindent}}


\def\CA{{\cal A}}

\def\CD{{\cal D}}
\def\CE{{\cal E}}
\def\CF{{\cal F}}

\def\CH{{\cal H}}

\def\CL{{\cal L}}

\def\CN{{\cal N}}
\def\CO{{\cal O}}

\def\CU{{\cal U}}

\def\CW{{\cal W}}

\def\p{\partial}

\def\ib{\bar{i}}

\def\lb{\bar{l}}


\def\Tr{{\rm Tr}}


\def\inbar{\,\vrule height1.5ex width.4pt depth0pt}

\font\cmss=cmss10 \font\cmsss=cmss10 at 7pt

\def\a{{\alpha}}
\def\ap{{\a}^{\prime}}
\def\b{{\beta}}
\def\d{{\delta}}
\def\g{{\gamma}}
\def\e{{\epsilon}}
\def\z{{\zeta}}
\def\ve{{\varepsilon}}
\def\vf{{\varphi}}
\def\m{{\mu}}
\def\n{{\nu}}

\def\l{{\lambda}}
\def\s{{\sigma}}
\def\t{{\theta}}
\def\o{{\omega}}
\def\nc{noncommutative\ }

\def\hp{\hat\partial}
\def\kk{{\kappa}}
\def\lref{\begingroup\obeylines\lr@f}
\def\lr@f#1#2{\gdef#1{\ref#1{#2}}\endgroup\unskip}

\lref\ikkt{N.~Ishibashi, H.~Kawai, Y.~Kitazawa, and A.~Tsuchiya,
\np{498}{1997}{467}, hep-th/9612115}
\lref\mtoda{P.~Etingof,
I.~Gelfand, V.~Retakh, ``Factorization of differential operators,
quasideterminants, and nonabelian Toda field equations''
q-alg/9701008} \lref\curtjuan{C.~G.~Callan, Jr., J.~M.~Maldacena,
\np{513}{1998}{198-212}, hep-th/9708147} \lref\bak{D.~Bak,
\pl{471}{1999}{149-154}, hep-th/9910135} \lref\baklee{D.~Bak,
K.~Lee, hep-th/0007107} \lref\moriyama{S.~Moriyama,
hep-th/0003231} \lref\wadia{A.~Dhar, G.~Mandal and S.~R.~Wadia,
\mpl{A7}{1992}{3129-3146}\semi A.~Dhar, G.~Mandal and S.~R.~Wadia,
\ijmp{A8}{1993}{3811-3828}\semi A.~Dhar, G.~Mandal and
S.~R.~Wadia, \mpl{A8}{1993}{3557-3568}\semi A.~Dhar, G.~Mandal and
S.~R.~Wadia, \pl{329}{1994}{15-26}}

\lref\mateos{D.~Mateos, ``Noncommutative vs. commutative descriptions of
D-brane BIons'', hep-th/0002020}
\lref\mrs{S.~Minwala, M.~ van Raamsdonk, N.~Seiberg,
``Noncommutative Perturbative Dynamics'', hep-th/9912072}
\lref\nahm{W.~Nahm, \pl{90}{1980}{413}\semi
W.~Nahm, ``The Construction of All Self-Dual Multimonopoles
by the ADHM Method'', in ``Monopoles in quantum field theory'', Craigie et
al., Eds., World Scientific, Singapore (1982) \semi
N.J.~Hitchin, \cmp{89}{1983}{145}}
\lref\rs{M.~van Raamsdonk, N.~Seiberg, ``Comments of Noncommutative
Perturbative Dynamics'', hep-th/0002186, \jhep{0003}{2000}{035}}
\lref\k{M.~Kontsevich, ``Deformation quantization of Poisson
manifolds'', q-alg/9709040}
\lref\gms{R.~Gopakumar, S.~Minwala, A.~Strominger,
hep-th/0003160, \jhep{0005}{2000}{020}}
\lref\agms{M.~Aganagic, R.~Gopakumar, S.~Minwala, A.~Strominger,
hep-th/0009142 }
\lref\sst{N.~Seiberg, L.~Susskind, N.~Toumbas, hep-th/0005040}
\lref\sdual{R.~Gopakumar, S.~Minwala, J.~Maldacena, A.~Strominger,
hep-th/0005048\semi
O.~Ganor, G.~Rajesh, S.~Sethi, hep-th/00050046}
\lref\filk{T.~Filk, ``Divergencies in a Field Theory on  Quantum Space'',
\pl{376}{1996}{53}}
\lref\cf{A.~Cattaneo, G.~Felder, ``A Path Integral Approach to the Konstevich
Quantization Formula'', math.QA/9902090}

\lref\cds{A.~Connes, M.~Douglas, A.~Schwarz,
\jhep{9802}{1998}{003}} \lref\douglashull{M.~Douglas, C.~Hull,
``D-Branes and the noncommutative torus'', \jhep{9802}{1998}{008},
hep-th/9711165}
\lref\wtnc{E.~Witten, \np{268}{1986}{253}}
\lref\volker{V.~Schomerus, ``D-Branes and Deformation
Quantization'', \jhep{9906}{1999}{030}}
\lref\cg{E.~Corrigan,
P.~Goddard, ``Construction of instanton and monopole solutions and
reciprocity'', \anp {154}{1984}{253}}

\lref\donaldson{S.K.~Donaldson, ``Instantons and Geometric
Invariant Theory", \cmp{93}{1984}{453-460}}

\lref\nakajima{H.~Nakajima, ``Lectures on Hilbert Schemes of
Points on Surfaces''\semi AMS University Lecture Series, 1999,
ISBN 0-8218-1956-9. }

\lref\neksch{N.~Nekrasov, A.~S.~Schwarz, hep-th/9802068,
\cmp{198}{1998}{689}}

\lref\freck{A.~Losev, N.~Nekrasov, S.~Shatashvili, ``The Freckled
Instantons'', {\tt hep-th/9908204}, Y.~Golfand Memorial Volume,
M.~Shifman Eds., World Scientific, Singapore, in press}

\lref\rkh{N.J.~Hitchin, A.~Karlhede, U.~Lindstrom, and M.~Rocek,
\cmp{108}{1987}{535}}

\lref\alexios{A.~Polychronakos, ``Flux tube solutions in
noncommutative gauge theories'',
hep-th/0007043}
\lref\branek{H.~Braden, N.~Nekrasov, hep-th/9912019\semi
K.~Furuuchi, hep-th/9912047}
\lref\davakisun{D.~Gross, A.~Hashimoto, N.~Itzhaki, ``Observables in
Non-commutative Gauge Theories'', hep-th/0008075}
\lref\wilsloops{N.~Ishibashi, S.~Iso, H.~Kawai, Y.~Kitazawa, hep-th/9910004,
\np{573}{2000}{573-593} \semi
J.~Ambjorn, Y.M.~Makeenko, J.~Nishimura, R.J.~Szabo, hep-th/9911041,
\jhep{9911}{1999}{029} ; hep-th/0002158, \pl{480}{2000}{399-408};
hep-th/0004147, \jhep{0005}{2000}{023}}

\lref\wilson{G.~ Wilson, ``Collisions of Calogero-Moser particles
and adelic Grassmannian", Invent. Math. 133 (1998) 1-41.}

\lref\gkp{S.~Gukov, I.~Klebanov, A.~Polyakov,
hep-th/9711112, \pl{423}{1998}{64-70}}
\lref\abs{O.~Aharony, M.~Berkooz, N.~Seiberg,
hep-th/9712117, \atmp{2}{1998}{119-153}}

\lref\abkss{O.~Aharony, M.~Berkooz, S.~Kachru, N.~Seiberg,
E.~Silverstein, hep-th/9707079, \atmp{1}{1998}{148-157}}

\lref\witsei{N.~Seiberg, E.~Witten, hep-th/9908142, \jhep{9909}{1999}{032}}
\lref\kinks{E.~Teo, C.~Ting, ``Monopoles, vortices and kinks in the
framework of noncommutative geometry'',
\pr{D56}{1997}{2291-2302}, hep-th/9706101}
\lref\kkn{V.~Kazakov, I.~Kostov, N.~Nekrasov,
``D-particles, Matrix Integrals and KP hierachy'', \np{557}{1999}{413-442},
hep-th/9810035}
\lref\tdgt{papers on 2d YM}
\lref\manuel{D.-E.~Diaconescu, \np{503}{1997}{220-238}, hep-th/9608163}

\lref\genmnp{L.~Jiang,
``Dirac Monopole in Non-Commutative Space'', hep-th/0001073}
\lref\hashimoto{K.~Hashimoto, H.~Hata, S.~Moriyama,
hep-th/9910196, \jhep{9912}{1999}{021}\semi
A.~Hashimoto,
K.~Hashimoto, hep-th/9909202, \jhep{9911}{1999}{005}\semi
K.~Hashimoto, T.~Hirayama, hep-th/0002090}
\lref\hklm{J.~Harvey, P.~Kraus, F.~Larsen, E.~Martinec,
hep-th/0005031}

\lref\snyder{H.~S.~Snyder, ``Quantized Space-Time'', \pr{71}{1947}{38};
``The Electromagnetic Field in Quantized Space-Time'', \pr{72}{1947}{68}}
\lref\connes{A.~Connes, ``Noncommutative geometry'', Academic Press (1994)}
\lref\planar{A.~Gonzalez-Arroyo,  C.P.~Korthals Altes,
``Reduced model for large $N$ continuum field theories'', \pl{131}{396}{1983}}
\lref\barsminic{I.~Bars, D.~Minic, ``Non-Commutative Geometry on a
Discrete Periodic Lattice and Gauge Theory'', hep-th/9910091}
\lref\grossneki{D.~Gross, N.~Nekrasov, ``Monopoles and Strings in
Noncommutative Gauge Theory'', \jhep{0007}{2000}{034},
hep-th/0005204} \lref\grossnekii{D.~Gross, N.~Nekrasov, ``Dynamics
of Strings in Noncommutative Gauge Theory'', hep-th/9907204}

\Title{\vbox{\baselineskip 10pt \hbox{PUPT-1961}
\hbox{ITEP-TH-58/00} \hbox{NSF-ITP-00-116} \hbox{hep-th/0010090}
{\hbox{ }}}} {\vbox{\vskip -30 true pt \centerline{
SOLITONS IN}
\smallskip
\smallskip
    \centerline{NONCOMMUTATIVE GAUGE THEORY}
\medskip
\vskip4pt }} \vskip -20 true pt \centerline{ David J.~Gross
$^{1}$, Nikita A.~Nekrasov $^{2}$}
\smallskip\smallskip
\centerline{ $^{1}$ \it Institute for Theoretical Physics,
University of California Santa Barbara, CA 93106 USA}
\centerline{$^{2}$ \it Institute for Theoretical and Experimental
Physics, 117259 Moscow, Russia} \centerline{$^{2}$ \it Joseph
Henry Laboratories, Princeton University, Princeton, New Jersey
08544 USA} \centerline{$^{2}$ \it Institut des Hautes Etudes
Scientifiques, Le Bois-Marie, Bures-sur-Yvette, F-91440 France}
\medskip \centerline{\rm e-mail: gross@itp.ucsb.edu,
nikita@feynman.princeton.edu}
\bigskip

\centerline{\bf Abstract} \medskip
\noindent

We present a unified treatment of classical solutions of
\nc gauge theories.
We find {\it all} solutions of the \nc Yang-Mills equations of motion
in 2 dimensions; and show that they
are labeled by two integers---the rank of the gauge group
   and the magnetic charge. The magnetic vortex solutions
   are unstable in 2+1 dimensions,
but correspond to the   full, stable BPS solutions of ${\CN} =4$ $U(1)$
\nc gauge theory in 4
dimensions, that  describes  $N$ infinite D1
strings that pierce a D3 brane at various points, in the presence of a
background $B$-field in the Seiberg-Witten ${\ap} \to 0$ limit. We discuss the
behavior of gauge invariant observables in the background
of the solitons.
We  use these solutions   to construct a panoply of
BPS and non-BPS solutions of supersymmetric gauge theories
that describe various configurations of D-branes. We analyze the instabilites of
the non-BPS solitons.
We also present an exact analytic solution  of
\nc gauge theory that describes a  $U(2)$ monopole.

\hfill\eject
\newsec{Introduction}
Recently there has been much interest in the properties of \nc
gauge theories.  The interest in these theories was sparked by the
discovery that these emerge as limits of M-theory
compactifications \cds\ or of string theory  with   D-branes in
the presence of a background Neveu-Schwarz $B$-field
\douglashull\volker\witsei,
   by
the  many analogies between
noncommutative gauge theories and
large $N$  non-abelian gauge theories \planar\filk\mrs, and
by  the many features that noncommutative field
theories share  with open string  theory \wtnc\mrs\rs.

In previous papers we  constructed and analyzed classical
solitons of \nc gauge theories  \grossneki,
   \grossnekii.  We first constructed exact, BPS, monopole solutions of
\nc $U(1)$ gauge theory. The solutions   were nonsingular and sourceless,
and described
smeared monopoles connected to a string-like
flux tube \grossneki. We interpreted this
string-monopole as the reflection of a D1 string attached to a
D3 brane in the presence of a background Neveu-Schwarz $B$-field,
in the decoupling Seiberg-Witten limit. The calculation of the tension
of the string, in precise agreement
with that expected from the D1 string, confirmed this picture.
In \grossnekii\ we constructed   an extremely
simple classical BPS solution of noncommutative $U(1)$ gauge theory
that describes an infinite D1 string piercing the D3 brane, which we
    called the{ \it fluxon }. (See also   \alexios)
   We were able to
evaluate explicitly the
complete spectrum of fluctuations about the
fluxon. We found   that the fluctuating modes are those of various strings,
connected to the D1 string and to the D3 brane.

In this paper we shall present a more unified description of
classical solutions of \nc gauge theory.  We will be
studying non commutative space of two dimensions,
namely a space whose coordinates satisfy $[x^i,x^j]=-i\t^{ij}$, where
the antisymmetric matrix $\t^{ij}$ has only two nonvanishing components,
   say $\t^{12}=-\t^{21}=\t$, although many of our considerations can be
easily generalized.

In Section 2 we discuss the properties of
\nc gauge theory. The standard procedure in constructing a \nc field
theory is to start
with an ordinary commutative field theory and replace ordinary products
with $\star$ products. In the case of gauge theories one starts with, say, a
   $U(N)$ gauge theory and replaces ordinary products
with $\star$ products in the definition of the field strength,
the gauge transformation and the action. Instead  we shall proceed
more abstractly.
The resulting \nc gauge theory  turns out to include $U(N)$ \nc gauge theory
{\it for all values of N} ! The value of $N$ will emerge as a superselection
parameter. labeling separate sectors
of the quantum Hilbert space. We discuss the gauge invariant observables
of this theory, the momentum carrying Wilson loops and current densities
of fields transforming in the fundamental representation of the gauge group.

In Section 3 we shall give a complete classification and
construction of all  co-dimension two classical solutions of \nc
gauge theory. These are solutions that are independent of time and
all but the two \nc spatial directions. They can describe
instantons of a Euclidean 2 dimensional \nc gauge theory, magnetic
vortex solitons  of a  2+1 dimensional \nc gauge theory, or vortex
string  solitons  of a  3+1 dimensional \nc gauge theory. Thus
they are  relevant for discussing D(p-2)-branes  attached, or
immersed in, Dp-branes. The fluxons that we previously considered
are special cases  of these solutions, but now we identify extra
moduli of these solutions that can be identified as the positions
of the vortices. We show that the gauge invariant observables of
the theory, when calculated in the soliton background, can be used
to measure these positions. We also show that, except in the case
where the solitons are BPS, the vortices are unstable and can
decay by spreading out in the \nc space. This is expected for D0
(or D1) branes immersed in D2 (or D3) branes.\foot{As this work
was being completed a discussion of these unstable solitons
appeared in \agms. } We  use these solutions   to construct a
panoply of BPS and non-BPS solutions of supersymmetric gauge
theories that describe various configurations of D-Branes. We
analyze the instabilites of the non-BPS solitons.

In Section 4  we give an explicit construction of a BPS monopole
in \nc  U(2) Higgsed gauge theory, by solving the \nc Nahm
equations. This is the \nc analogue of the t'Hooft-Polyakov
monopole and corresponds precisely, as we show, to a D1 string
stretched between two separated D3 branes in the Seiberg-Witten
decoupling limit. This is a localized soliton in three dimensions,
with an interesting internal  spatial structure.

\newsec{Noncommutative gauge theory}

\subsec{Noncommutative space }

Consider   2+1 dimensional   space-time   with coordinates
$x^i$, $i=1, 2 $ which obey the following commutation
relations:
\eqn\cmrl{[x^i, x^j] = -i {\t^{ij}} , \, \t^{12}=\t,\quad  [t,x^1]=[
t,x^2]=0\ . }
By noncommutative space-time we mean the algebra ${\CA}_{\t}$
generated by the $x^i$ satisfying \cmrl . We can think of elements of
the algebra
as functions of the operators $x^i$, together with some extra
conditions on the allowed expressions in the $x^i$. We shall largely
suppress the
dependence on the commutative coordinate $t$.

It is convenient to introduce the creation and annihilation operators:
\eqn\cs{ c^\dagger={1\over \sqrt{2\t}}(x^1+ix^2), \quad
   c ={1\over \sqrt{2\t}}(x^1-ix^2);  \quad [c ,c^\dagger]=1\ . }
The spatial coordinates can then be thought of as operators in the space
of Fock states:
\eqn\fock{ \CH = \{  \vert 0 \rangle, \vert 1 \rangle, \dots \vert n
\rangle, \dots   \}\ , }
where
\eqn\fockk{c\vert 0 \rangle=0, \quad \vert n \rangle=
{{c^\dagger}^n\over \sqrt{n!}}\vert 0 \rangle,
\quad {c^\dagger}{c }\vert n \rangle= n\vert n \rangle\ .  }
Elements of the algebra ${\CA}_{\t}$
are then represented as operators in this Fock space,
$f(c,c^\dagger)$.

The elements
of ${\CA}_{\t}$ can also be identified with ordinary functions on ${\bf
R}^2$, with the product of two functions $f$ and $g$ given by the
Moyal formula (or star product):
\eqn\myl{ f \star g \, (x)=
{\exp} \left[ {i \over 2} {\t}^{ij} { {\p} \over {\p x_{1}^{i} }}
{{\p}\over {\p x_{2}^{j}}} \right] f(x_{1}) g (x_{2}) \vert_{x_{1}
= x_{2} = x}\ .}
For  plane waves:
\eqn\mylw{ e^{i {\vec p}_1
{\cdot } {\vec x} } \star e^{i {\vec p}_2 \cdot {\vec x}} = e^{-
{i\over 2} {\vec p}_1 \times {\vec p}_2}\quad e^{i ({\vec p}_1 +
{\vec p}_2) \cdot {\vec x}}\ , }
where
\eqn\vpr{{\vec p}_1 \times
{\vec p}_2 = {\t}^{ij} p_{1 i} p_{2 j} = - {\vec p}_2 \times {\vec
p}_1 \ .}

The procedure that maps
ordinary commutative functions onto
operators in the Fock
space is called Weyl ordering and is
defined by:
\eqn\wlor{\eqalign{ & f(x)=f\left( z=x^1-
ix^2, \bar z=x^1+ix^2\right)
    \mapsto  \cr & {\hat f(c, c^\dagger)}
=  \int f(x) \, {{d^{2} x \,\, d^{2} p
}\over{(2{\pi})^{2}}} \,
\,  e^{ i \left[ {\bar p}_a \left( \sqrt{2\t} c  - z \right)
+ p_{a}
\left( \sqrt{2\t}{c^\dagger}  -
{\bar z} \right) \right]  } \ ,  }   }
where $p=(p^1+ip^2)/2, \, \bar p=(p^1-ip^2)/2$.
Conversely:
\eqn\converse{    f(z, \bar z) =  \pi \t \int  { d^{2} p  \over{(2{\pi})^{2}}}
      {\Tr} \left\{ \hat f(c, c^\dagger)  \, \,
\,  e^{ -i \left[ {\bar p}_a \left( \sqrt{2\t} c  - z \right)
+ p_{a}
\left( \sqrt{2\t}{c^\dagger}  -
{\bar z} \right) \right] }\right\}  \ . }
It is easy to see that
\eqn\product{  {\rm if } \quad f \mapsto \hat
f, \quad g \mapsto \hat g \quad {\rm then
}\quad f\star g \mapsto
\hat f \hat g \ .}
We also note that
\eqn\integ{\int d^2 x f(x) \mapsto \pi \t {\Tr} \hat f(c,c^\dagger) \ .}
The
derivative ${\p}_i$ is an infinitesimal
automorphism of the
algebra \cmrl:
\eqn\auto{x^i \to x^i +
{\ve}^i,}
where ${\ve}^i$ is a $c$-number. For the algebra
\cmrl\ this automorphism
is internal:
\eqn\intrn{{\p}_i {f} =  {{d\over d \e^i } \, f (x^i+\e^i\cdot
1)\vert}_{\e^i=0}
= \,  i  {\t}_{ij} [
    x^j,{f}] = \, [ {\hp}_i, f ] ,}
where ${\t}_{ij}$ is the inverse of
${\t}^{ij}$, namely
${\t}_{ij}{\t}^{jk}=\delta^k_i$, and ${\hp}_i = i{\t}_{ij}x^{j}$.
Thus translations in the Fock space are
generated by $\hp_i = i  {\t}_{ij}x^j$,
so that  if $ f(x) \mapsto \hat f$, then $ f(x
+a) \mapsto \exp(a\cdot \hp)\hat f\exp(-
a\cdot \hp)$.

\subsec{Gauge theory }

The standard procedure in constructing a \nc field theory is to start
with an ordinary commutative field theory and replace ordinary products
with $\star$ products. In the case of gauge theories one starts with, say, a
   $U(N)$ gauge theory and replaces ordinary products
with $\star$ products in the definition of the field strength,
the gauge transformation and the action. Here we shall proceed more abstractly.
The resulting \nc gauge theory  turns out to include $U(N)$ \nc gauge theory
{\it for all values of N}.

Gauge fields arise most naturally  via covariant derivatives.  In other words,
we first consider matter fields, $\Psi$, which form a  representation
of the gauge group,
or  of the gauge algebra,
and then form  a covariant derivative, $\nabla_i $, so that $\nabla_i \Psi$
is a matter field in the same representation.

The abstract definition of matter fields on a \nc space is simply that they are
representations of the algebra of \nc functions, ${\CA}_{\t}$. Thus
if $f$ is an element of the algebra
then $\Psi$ is  a matter field if
\eqn\matter{f:\Psi \mapsto f\cdot \Psi, \quad {\rm where} \, \,
f\cdot (g\cdot \Psi)
=(f\star g)\cdot \Psi \ .}
(strictly speaking \matter\ covers only the so-called left modules,
there are also right modules, for which:
$f \cdot ( g \cdot \Psi) = (g \star f) \cdot \Psi$).
Given such a representation we
then search for a covariant derivative
$\nabla_i$ that satisfies the Leibnitz rule:
\eqn\Leib{\nabla_i(f\cdot \Psi)= (\partial_i f)\cdot \Psi +
f\nabla_i\cdot \Psi\ .}
Since the derivative of $f$ satisfies the Leibnitz rule
$$ \p_i (f \star g) =[\hp_i, f \star g]=[\hp_i,f]\star g+f\star [\hp_i,g]=
(\p_i f )\star g +f\star (\p_i g),
$$
this ensures that $\nabla_i\Psi$ is in the same representation of the algebra
${\CA}_{\t}$. (Again, this definition of the covariant derivative
is specific   to algebras like ${\CA}_{\t}$ which have enough
  translational symmetries. We do not need here the more general definition
of the
gauge field given by Connes\connes).

The simplest representation of the algebra, which is
equivalent to operators in the Fock space $\CH$, is the Fock space itself,
the Fock representation $\vert \Psi_F\rangle =
\sum_{n=0}^\infty\Psi_n\vert n \rangle $.
Clearly this  is a representation with $f:\Psi_F \mapsto f\vert \Psi_F\rangle$.
What are the possible covariant derivatives? If the Leibnitz rule is satisfied
then
\eqn\leb{[\nabla_i, f]\vert \Psi_F\rangle = i\t_{ij}[x^j,f]\vert \Psi_F\rangle
=\partial_if\vert \Psi_F\rangle\ . }
   Consequently
\eqn\conseq{\left[ \nabla_i-i\t_{ij} x^j,f\right]\vert \Psi_F\rangle=0 \ ,}
for any $f$ in the algebra and for any  $\vert \Psi_F\rangle$ in $\CH$.
The unique solution  is that {\it all gauge fields are of the form:}
\eqn\gauge{\nabla_i= i\t_{ij} x^j + \a_i{\rm I }= \hat \partial_i  +
\a_i{\rm I }\ ,}
where ${\rm I }$ is the identity operator and $\a_i$ are c-numbers.
If we define the connection, or gauge field, to be
$A_i=\nabla_i -\hp_i$
then this gauge field is given by the  trivial
$$A_i =  \a_i{\rm I }\ .
$$
The  gauge transformations must commute with the action of the algebra
in our representation. In our example then these gauge transformation
must be given by multiplication by $U$,
\eqn\gtps{\vert \Psi_F\rangle \to U\vert \Psi_F\rangle \ ,
}
where $U$ is a c-number.  If we demand that the gauge transformations
preserve the  norm  of $\vert \Psi_F\rangle$, then
$U=\exp[i\alpha]$.

The triviality of the gauge transformations and the gauge field follows
from the fact that the matter
field $\vert \Psi_F\rangle$ is the analogue of a field
with support only at one point in space.
To illustrate this point better let us think of the points on an ordinary
commutative
space $X$ as  the irreducible representations of the algebra of functions
on this space, $A_0$. These appear in the decomposition of the algebra
viewed as
its own representation:
$$
f(x) = \int_{X}\, {\rm d}y \, f(y) \, P_{x}(y) \ ,
$$
where $P_{x}(y) = {\d}(x - y)$ is   not strictly speaking an element
of the algebra of smooth functions, but can be approximated by smooth
functions.
This relation can be written also as:
$$
{\CA}_{0} = C^{\infty}(X) = \bigoplus_{x \in X} R_{x}\ ,
$$
where $R_{x}$ is the one-dimensional irreducible
representation of ${\CA}_{0}$:
$$
R_{x}(f) \cdot \Psi = f(x) \Psi  \ .
$$
In the same fashion:
$$
{\CA}_{\t} = \bigoplus_{n \in {\IZ}^{+}} {\CH}_{n} \ .
$$
where ${\CH}_{n}$ is a  representation of ${\CA}_{\t}$ isomorphic to ${\CH}$:
$$
f \in {\CA}_{\t}, \quad {\hat f}\vert n \rangle \in {\CH}_{n}\ .
$$
In this sense the Fock representation is the analogue of a single
point on the \nc space.
If we take  a direct sum of $k$ copies of
${\CH}$ as another example of representation of ${\CA}_{\t}$ then
the gauge fields will become $k \times k$ matrices $A_i$
and the gauge transformations will form the unitary group $U(k)$.

Although translations act
non-trivially on
$\vert \Psi_F\rangle$, as $\vert \Psi_F\rangle\to
\exp[a\cdot \hat \partial] \vert \Psi_F\rangle$, this can also be regarded
as   a gauge transformation,$\vert \Psi_F\rangle \to f \vert \Psi_F\rangle$,
with $f=\exp[ ia^i\t_{ij}x^j]$.

To construct a matter field defined over all of the \nc plane
we take
\eqn\psimat{\vert \Psi\rangle =\sum_{nm} \Psi_{nm}\vert n\rangle\langle m \vert
= \sum_m\left\{ \sum_n \Psi_{nm}\vert n\rangle\right \} \langle m \vert  \ ,  }
which is a infinite sum of Fock representations, one for each point,
$ \langle m \vert $, on the \nc plane.  In fact $\Psi$
is simply an element of the algebra itself,
an operator on the Fock space, and the representation is
$$f:\Psi \mapsto f\Psi  \quad {\rm or}\, \,  f:\Psi \mapsto \Psi f \ .
$$
   Let us consider the representation $ f:\Psi \mapsto \Psi f$, which
we shall call
the fundamental representation
(it is an example of a {\it right} \   module;
  whereas $\Psi^\dagger$   transforms by  multiplication on the left and
thus forms
a left ${\CA}_{\t}$-module).
Therefore a field in the fundamental representation, is represented
by the operator
    $\sum \psi_{n,m}\vert n\rangle\langle m\vert $, where
   $\vert n\rangle $ carry all the information about the $U(\infty )$ gauge
and  $\langle m\vert$ carry all the positional information
(vice-versa for $\Psi^\dagger$).

Now we  have more freedom in constructing the covariant derivative, in fact
\eqn\covder{\nabla_i \Psi = -i\Psi \t_{ij}x^j + D_i \Psi \ ,}
   will satisfy the Leibnitz rule, where $D_i$ is any anti-Hermitean
operator in the Fock space.
Since the ordinary derivative of $\Psi$ is
   $\partial_i\Psi =[\hat \partial_i,\Psi] =[ i \t_{ij}x^j, \Psi] $, we
shall write
\eqn\D{ D_i = i \t_{ij}x^j +A_i ; \quad
\nabla_i \Psi = [\hat \partial_i,\Psi]  + A_i \Psi \  .}

Gauge transformations, that preserve the representation, $ f:\Psi
\mapsto \Psi f  $,
of the algebra are given by
$$\Psi\to \Psi^U=U\Psi\ .
$$
To preserve the norm of $\Psi$ we demand that $ U^\dagger U={\rm I}$.
Consequently, $\Psi^\dagger \Psi$ is a
gauge invariant, local, observable.
Under such a gauge transformation the covariant derivative of $\Psi $ should
transform in the same way as $\Psi$, so that:
\eqn\covtr{D_i\Psi\to U D_i\Psi= (UD_iU^\dagger)U\Psi\ . }
Consequently under gauge transformations:
\eqn\gt{ D_i\to UD_iU^\dagger \, ; \quad A_i\to U [\hat
\partial_i,U^\dagger]   +
UA_iU^\dagger \ . }
We shall define the field strength, $F_{ij}$, as usual,
\eqn\field{F_{ij} = [\nabla_i,\nabla_j]=[D_i,D_j]-i\t_{ij}\ .}

The covariant derivative $D_i$ transforms  in the adjoint representation,
like a matter field $\Phi$ (an element of the algebra, an operator in
the Fock space), whose covariant derivative is
$$\nabla_i \Phi=[D_i,\Phi]\ .$$
   Under translations both $D_i$ and $\Phi$ transform under
a subgroup of the gauge group:
$$(\Phi,D_i) \to\exp[ia^i\t_{ij}x^j](\Phi,D_i)\exp[-ia^i\t_{ij}x^j]\ , $$
consequently gauge invariant observables constructed from these fields
will be translationally invariant.

A gauge invariant bosonic action, or since we supressing the time dependence,
an energy density $\CE$,
can then be formed as
\eqn\acttt{\CE = {\Tr}\left\{\left([D_i,D_j]-i\t_{ij}\right)^2
+ \sum_a \left( [D_i,\Phi_a]^2 + V(\Phi_a)\right)
+ \Psi^\dagger_a(D_i^2+m^2_a)\Psi_a\right\} \ .}

The gauge group  under which this density is invariant, generated by  \gtps\
and  \gt, is that of unitary operators in $\CH$, or $U(\infty)$. Nonetheless,
we shall see  that, regarded as a functional of $D_i$, $\Phi_a$ and
$\Psi_a$, this
expression contains all possible \nc gauge theories with gauge group $U(N)$,
   for all $N=0,1,2,\dots \ .$ To see this let us ignore the matter fields
and write the action for the 2+1
dimensional \nc     gauge theory, in the gauge $A_t = 0$, as:
   \eqn\tpoo{S = {{2\pi \t}\over{4g^2}} \int {\rm d}t {\Tr}\,
\left\{\partial_tD \partial_t \bar D
-4 \left[ [\bar D,   D] +{1\over 2\t}\right]^2 \right\}\ .}
   where we have rewritten the covariant derivatives as:
\eqn\opr{D =  - {c^{\dagger}\over \sqrt{2\t}} + {\half}\left( A_1 + i
A_2 \right),
\quad
{\bar D} = {c\over \sqrt{2\t}} + {\half}\left( A_1 - i A_2 \right) =
- D^{\dagger}\ ,  }
so that the  field strength is given by:
\eqn\field{F=F_{1,2} =  2\left[ [\bar D,  D] +{1\over 2\t} \right]  \ .}

   The physics of this system is  given by   the  infinite-dimensional
space ${\CF}$ of the operators $D, {\bar D}$ acting in the Fock space ${\CH}$
moded out by the action of the gauge group ${\CU}$ of unitary
operators in ${\CH}$, acting via:
\eqn\gtr{
D \mapsto U D U^{\dagger}, \quad \bar D \mapsto U \bar D U^{\dagger},
\quad U U^{\dagger}  = U^{\dagger} U = 1\ . }
We now argue that this quantum mechanical system describes $U(N)$ \nc
gauge theory for all values of $N$, where $N$ is a superselection
parameter. The argument is that for the energy of a field
configuration to be finite,
or for the action to be finite, $F$ must vanish almost everywhere---i.e.
except for a finite number of matrix elements we must have that
$$[  D^\dagger  ,  D] ={1\over 2\t}\ .$$
This is obviously true of the absolute minima of the action----the
classical vacua.
The unique irreducible represenation  of   this Heisenberg
algebra  is, up to unitary equivalence,$$D=-c^{\dagger}/\sqrt{2\t},\,
D^\dagger= -c/\sqrt{2\t}\  ,$$
and the most general representation, classical vacuum, is a reducible
sum of $N$ such
irreducible representations, acting in the direct product of $N$
copies of $\CH$,
$\CH=\CH\oplus \CH\oplus \dots \CH \approx {\CH} \otimes {\IC}^{N},$
which is isomorphic to $\CH$ itself (by a version of the ``Hilbert hotel''
argument).
We label the  basis vectors of this space
by $\vert n,a\rangle, a=1\dots N $, and in this basis:
\eqn\DN{D^{(N)} = -c^\dagger/\sqrt{2\t}\otimes {\rm I}_{N}, \, \,{\rm
where} \, \, c^\dagger \vert n,a\rangle
=\sqrt{n+1}\vert n+1,a\rangle,\,\, {\rm I}_{N}
\vert n,a\rangle=\vert n,a\rangle . }
This vacuum is invariant under a  $U(N) $ gauge
transformations that act on the $a$ labels. $N$ is an index. It is equal to the
difference of the number of zero eigenvalues of the Hermitean operators
$  D^\dagger  D$ and  $D D^\dagger  $, whose non-zero eigenvalues coincide.

As far as we can ascertain the quantum theory constructed about one
of these vacua
will not mix with the others. Any path in field space that connects
different vacua has infinite energy and action.  Thus the functional integral
for the partition function with action given by \tpoo\
   breaks up into a sum of partition  functions for each $U(N)$ gauge theory:
\eqn\funct{Z=\int {\CD D \CD \bar D\over {\rm Vol}
U(\infty)}\, \, \exp[iS]=\sum_{N=0}^\infty Z_N \ .}

In the sector labeled
by $N$ we would expand
   $D=D^{(N)}+A^{(N)}\ , $ and
would find that the field strength takes the  customary form
for the  field strength of a $U(N)$ \nc  gauge theory:
$$
F={1\over\sqrt{2\t}}\left(  [c\cdot I,{\bf A} ] +
[c^\dagger\cdot I,{\bf \bar A}] \right) + [{\bf \bar A},
{\bf A}] \ ,
$$
where
   ${\bf A}_{ab}$
is an $N\times N$ matrix
operator  in  ordinary Fock space, given in terms of $A^{(N)}$  as:
$$ \langle m\vert {\bf A}_{ab}\vert n\rangle =
\langle m,a\vert A^{(N)}\vert n,b\rangle \ . $$
It is fascinating
that the action for \nc gauge theory
does not determine the rank
of the gauge group, but rather that it emerges as a superselection parameter.

\subsec{Gauge invariant observables}

What are the physical observables of this theory?
They, of course, should be invariant under the
$U(\infty )$ gauge transformations \gtr.
Since  translation of the \nc coordinates are generated, up to shifts of
the gauge field, by
these unitary transformations, there appears to a conflict between
gauge invariance and locality. The simplest gauge invariant observables are
in fact  non-local.

For any $l  \in {\IC}$, $l = l_1  + i l_2$, consider the operator
$$ {\CD} (l) = {\lb} D + l D^{\dagger} = l_1D_1+l_2D_2 \ . $$ This
is a Hermitean operator whose eigenvalues are   gauge invariant
functions on ${\CF}$. Consequently,  traces of ordered products of
exponentials of $i {\CD} (l_a)$ for different $l_a$: \eqn\wlsn{
{\CW} ({\vec l}) = {\Tr}\ {\prod}_{a} {\exp} i{\CD} (l_a)\ , }
provide  a set of gauge-invariant functions on ${\CF}$.  The
functionals \wlsn\ are the \nc analogues of the familiar Wilson
loops \wilsloops\davakisun, that describe parallel transport along
the path described by the series of displacements along $l_a$.
Unlike the commutative case
    the ``path'' along which the loop is taken does not need
to   be closed: $\sum_a l_a \neq 0$ in general, and ${\CW} ({\vec l})$
is not a local operator---rather it has momentum equal
to $\theta^{-1}_{ij}\sum_a l_a^j.$

Another set of gauge-invariant observables can be constructed in terms of the
operators $D$ and $\bar D$, or in terms of bilinears in fields that transform
in the fundamental representation.  Consider the
space of normalizable
solutions of the massless Dirac equation in the background gauge field
$A_1, A_2$.
We take the  fermions to transform  in the fundamental representation of
the gauge group,
\eqn\fermg{\pmatrix{ \psi_{L} \cr \psi_{R} } \mapsto
U \pmatrix{ \psi_{L} \cr \psi_{R} }\ . }
Then the Dirac equations are, in the operator formalism:
\eqn\drc{D {\psi}_{L}^{i} + {\psi}_{L}^{i} c^{\dagger} = 0, \quad
{\bar D} {\psi}_{R}^{j} - {\psi}_{R}^{j} c = 0}
where $i = 1, \ldots, n_{L}$ and $j = 1, \ldots, n_{R}$
and $n_L, n_R$ are the numbers of the left-moving and the right-moving
zero modes.
These equations are clearly invariant under  \fermg\ .
Under translations they transform as
\eqn\transla{{\psi}_{R,L}  \to  e^{a\cdot \hat
\partial}\psi_{R,L}e^{-a\cdot \hat \partial}
\ .}
As we noted before,  translations are  equivalent, up to gauge transformations,
   to transformations  of the type ${\psi}_{R,L}  \to
\psi_{R,L}e^{-a\cdot \hat \partial} $.

The components of the $U(n_L)_{L} \times U(n_{R})_{R}$ current:
\eqn\crrnt{j_{L}  = {\psi}_{L}^{\dagger}   {\psi}_{L} ,
\quad j_{R}  = {\psi}_{R}^{\dagger} {\psi}_{R} \ , }
as well as of the `density'
\eqn\dnsty{{\rho}  = {\psi}_{L}^{\dagger  } {\psi}_{R} \  , }
are gauge invariant observables of  the \nc gauge theory.
Unlike the Wilson loops these are {\it local} observables.
Under translations the transform as
$O\to e^{a\cdot \hat \partial}Oe^{-a\cdot \hat \partial}  $,
where $O$ stands for $j_{R},j_{L}$ or ${\rho} $.

\newsec{Classical static solutions}
In \grossnekii\ we constructed classical solutions of  \nc gauge
theories, fluxons. In
the supersymmetric
   3+1 \nc Yang-Mills theory gauge theory these described
D1 strings that pierced  a D3 brane and were BPS solutions. We also pointed
out that
   vortex line  solutions of 3+1 \nc Yang-Mills theory gauge theory or
point vortex
solutions of 2+1 \nc Yang-Mills theory could be easily  generated by
setting the scalar
field, $\Phi$,
that described the extension of the D1 string into the bulk, to zero.
We shall now construct all static solutions
of   pure 2+1 \nc Yang-Mills theory with finite energy.
We shall recover the  analogue of the N-fluxon solution, however with
additional
moduli, that we will see describe the position of the vortices.
  From now on we shall often  set $2 \t=1$, to simplify the formulae. $\t$
can always be
reintroduced by scaling the \nc coordinates as $x^i\to \sqrt{2\t} x^i$.

For time independent gauge fields the equations of motion are
\eqn\eqm{[D, [\bar D, D]]=[\bar D, [\bar D, D]]=0 \ ,}
and the energy is proportional to $ {\Tr} \left( [   D^\dagger, D  ]
- 1 \right)^2.$

\subsec{Classification of solutions}
Consequently we need
to find a pair of operators $D, D^{\dagger}=-\bar D$, acting in
${\CH}$, that obey:
\eqn\eom{\eqalign{& [   D^{\dagger}, D  ] =  1 +F \ ,\cr
& [ D, F ] = [ D^{\dagger}, F ] = 0,\cr
& {\Tr} F^2 \  < \  \infty \ .\cr}}
The first three equations imply that  $D, D^{\dagger}, 1+F$ form a
Heisenberg algebra,
with $F$ generating the  center of the algebra.
The Hilbert space ${\CH}$ decomposes into   irreducible representations of
this algebra,
with $F$ equal to  a constant
$f_n$ on the $n$'th component. Let $d_n$ be the dimension of the
$n$'th irreducible component. It is well-known that unless
$1 +f_n =0$ it must be that $d_n = \infty$.
The finite energy condition implies
that:
\eqn\enrg{\sum_n d_n f_n^2 < \infty \ . }
Therefore there are just two possibilities:
\eqn\pos{ f_n = 0, d_n = \infty,  \quad {\rm or} \quad  f_n =-1, d_n \geq
0  \ .}
Hence, by a unitary gauge transformation, we can bring $D$ and
$D^{\dagger}$ to the following form: on a finite dimensional subspace $V_q$
of dimensionality
$q$:
$$[ D, D^{\dagger}] = 0, \quad D = {\rm diag}\left( -{\l}_1, \ldots,- {\l}_q
\right)\ , $$
(We have chosen this sign convention so that, as we will see below, $\l_i$ will
be the position of the i$^{\rm th}$ vortex, for $2\t=2$ )
   while on the complement, ${\CH} \ominus V_q$, which is isomorphic to ${\CH}$,
$D$ is a reducible sum of $N$ irreducible representations of the
Heisenberg algebra
   ($D = - c^{\dagger}, \quad {\bar D} = c\  $),  as given explicitly in \DN\ .

Let us, for simplicity, choose $N=1$,
and let ${S_q}^{\dagger}: {\CH} {\ominus} V_q \to {\CH}$ be the unitary
isomorphism between the
two Hilbert spaces. We can extend ${S_q}^{\dagger} $ to the whole of
${\CH}$ by
having it act as $0$ on $V_q$.
Then $ {S_q}^{\dagger} $, as an operator in ${\CH}$, obeys:
\eqn\rlprj{{S_q}^{\dagger} {S_q} = 1, \quad S_q{S_q}^{\dagger} = 1 - P_{q}\ , }
where $P_{q}: {\CH} \to V_{q}$ is the orthogonal projection.
Again, by unitary gauge transformation we can assume that
$V_{q}$ is spanned by the vectors
$\vert 0 \rangle, \ldots, \vert q-1 \rangle$.
Thus, for $N=1$,  the generic static solution of the \nc Yang-Mills
equations of motion
is given by:
\eqn\gnrst{D = {\l}_{q} - S_q c^{\dagger} {S_q}^{\dagger}\ ,}
where $ {\bf \l_q}$ $ = \sum_{i=0}^{q-1} {\l}_i \vert i
\rangle\langle i \vert$,
and  $ S_q |n\rangle = |n+q\rangle  $.
This gauge field has   field strength:
\eqn\fld{F =   -P_{q}\ , }
which implies that the solution has a magnetic charge $q$.
The moduli $ {\bf \l_i}$ describe the positions of the vortices,
as we shall explicitly see below, by examining the behavior of the Wilson
loop or the
position dependent
fermion bilinears in this background.

Thus all classical static solutions of 2+1 dimensional \nc gauge theory
are classified by the rank of the gauge group, $N$, and by the magnetic charge,
$q$.
It is at first a little bit surprising to discover solutions of positive
charge, and to have no solutions of  negative charge. However,
the noncommutativity
breaks the left-right symmetry and as a consequence, one cannot
simply by change of orientation  produce   anti-vortices  from
vortices.

It is also surprising to find  that there is a $2q$-dimensional moduli
space of solutions for magnetic charge $q$  corresponding to
the separations of the vortices. Since the vortices are not BPS
solutions we would have expected them to repel. Indeed in a
commutative gauge theory we could have considered two vortices
very far way from each other, and then the one gauge boson
exchange interaction could have been calculated exactly----leading to
repulsion.
However,  in the \nc gauge theory the energy density is not gauge invariant
and indeed the energy density of the vortices, proportional to $
{\Tr} \left( [ D, D^{\dagger} ]
   +1 \right)^2$,
is independent of   $ {\bf \l_i}$, and no similar conclusion can be drawn.

\subsec{Fermion Condensates}
Let us now consider the behavior of fermions,
transforming in the fundamental representation,
in the presence of the multi-vortex solutions.
As we will see below, fermion bilinear  observables, such as \crrnt,
\dnsty,will
be sensitive to the moduli $\lambda_i$, i.e. to the position of the vortices.

The operator $D$ acting in ${\CH}$, being close to $ -c^\dagger $
has a finite spectrum of normalizable eigenstates.
Let $-{\e}_i$, $i=1, \ldots, n_L$ be the corresponding eigenvalues,
and ${\chi}_i \in {\CH}$ the corresponding eigenstates.
Then ${\psi}^i_{L} =  {\chi}_i \otimes \langle {\e}_i  |$
is a Dirac zeromode, with $\langle {\e}_i \vert$ being the coherent state
$$
\langle {\e}_i \vert \equiv  \langle 0 \vert e^{{\e}_i c}e^{-{1\over
2} \bar \e \e} \ .
$$
Then the corresponding current $j_L$ is diagonal and has the components:
$$
j_L^{{\ib}i} = \vert {\bar\e}_i \rangle \langle {\e}_i \vert \ .
$$

   On the other hand if we examine the  right handed modes ${\psi}^i_{R} $,
we will not find zero modes at all, since the analogue of
$\langle {\e}_i \vert $ would be a state $\langle {\s} \vert $
that would be an eigenstate of $c$, i.e.
$\langle {\s} \vert c=\e \langle {\s}\vert  $,
and there are no such normalizable states.

Consider the Dirac equation in the background of the q-vortex
solution \gnrst. We first look for eigenvectors of $D$, ${\chi}_i \in {\CH}$,
satisfying
$$D\, {\chi}_i\, = \, \left[
-{\l}_q - S_qc^\dagger {S_q}^\dagger \right] \, {\chi}_i \, = \, \e_i
   {\chi}_i \  .
$$
It is clear that there exist $q$ (and no more) solutions of this equation,
where
\eqn\thed{ \e_i= -\l_i, \quad \chi_i =|i\rangle,\quad i=0,\dots, q-1\ .}
So $n_L=q$, and
\eqn\zero{\Psi_L^i = \vert i \rangle \otimes \langle {\l}_i\vert   ,
\quad i=0, \dots ,
n_L-1
\ .}
The gauge invariant current is then given by:
\eqn\current{j_L^{{\ib}i} = q\vert {\bar \l}_i \rangle \langle{\l}_i \vert .}
Under translations this current transforms non trivially, in fact
under a spatial
translation by an amount $\Delta_i $, with $\Delta=\Delta_1+i\Delta_2$  ,
\eqn\trn{j_L^{{\ib}i}= q\vert {\bar \l}_i \rangle \langle{\l}_i \vert
\to e^{-c\bar \Delta   +c^\dagger \Delta  j_L^{{\ib}i} }
e^{-c\bar \Delta +c^\dagger \Delta }=q\vert {\bar \l}_i+\bar
\Delta \rangle \langle{\l}_i+ \Delta    \vert \ .}
Therefore we see that the $  \lambda_i $ can be interpreted as
the positions of the vortices. If we restore $\t$ we would find that
the positions $x^m_i$ of the vortices are gven by $x^m_i=\t^{\m
\n}\lambda_{i \n}$ (D
and thus $\lambda_i$ have
   the dimensions of momenta.) If we use \converse\ to write the
current as an ordianry function
we find that
\eqn\curfn{j_L^{{\ib}i}(x) \propto\exp\left[-(x-x_i)^2\over 2\t\right]\ ,}
  curresponding to a current density   that is a Gaussian
localized at position
$x_i $, of width proportional to $\sqrt\t$.

\subsec{Probing the vortices with Wilson loops}

The above discussion makes it clear that the moduli $\lambda_i$
correspond to the position of the vortices.
The positions of the
vortices can also be detected using the Wilson loop  operators of
definite momenta.
Let us evaluate the  operators \wlsn\ that correspond to a Wilson loop
of definite momentum in the background of the q-vortex
solution \gnrst\ . We reintroduce ${\t}$ in this
section. The trace separates as
\eqn\wlsnbk{{\CW} ({\vec l})= {\Tr}\ {\prod}_{j} {\exp} i {\CD} (l_j)
= \sum_{i=0}^{q-1} e^{i {\vec\ell} \cdot {\vec\l}_i}+ {\Tr}\ {\prod}_{j}
{\exp}{i\over\sqrt{2\t}}
\left[ S_q ({\lb}_{j} c^\dagger + l_{j}  c){S_q}^\dagger \right ]\ ,}
where ${\vec\ell} = \sum_j {\vec l}_j$.
   The  second term in \wlsnbk\  is  actually identical to the Wilson
loop \wlsn\ in the vacuum,
since using   ${S_q}^{\dagger} {S_q} = 1, \ \  S_q{S_q}^{\dagger} = 1
- P_{q}$, we have
\eqn\wltr{\eqalign{& {\Tr}\ {\prod}_{j}
{\exp}{i\over\sqrt{2\t}}\left[ S_q ({\lb}_{j} c^\dagger + l_{j}
c){S_q}^\dagger \right ]=
{\Tr}\  S_q \left[{\prod}_{j}
{\exp}  {i\over\sqrt{2\t}}({\lb}_{j} c^\dagger + l_{j}  c)\right]
{S_q}^\dagger  = \cr
&{\Tr}\   {\prod}_{j}
{\exp} {i\over\sqrt{2\t}}({\lb}_{j} c^\dagger + l_{j}  c) \quad =
\quad {\delta^2({\ell})\over 2\t}\exp\left[{{i\CA}\over{2\t}}\right]\ .} }
Where $\CA $ is the (oriented) area
spanned by the loop (delta function makes the loop close),
$\CA= {1\over{2i}} \sum_{i<j} \left(
l_i {\bar l}_j - l_j {\bar l}_i \right) =
{\half}\oint x_{1} dx_{2} - x_{2} dx_{1} $.

Thus, in  the leading semi-classical
approximation
$$\langle {\CW} ({\vec l})\rangle =
\sum_{i=0}^{q-1} e^{i{\vec \ell}\cdot {\vec \l}_i}+{{\d}^2( {\ell} )\over 2\t}
\exp[{{i\CA}\over{2\t}}]\ .
$$
Consequently ${\CW} ({\vec l})$, for nonvanishing momentum $p_{\m} =
{\t}^{-1}_{\m\n} {\ell}^{\n}$,  couples
to the vortices which behave as local, pointlike sources at the
positions $x^{\m} = {\t}^{\m\n}{\l}_{\n}$.
   By measuring  ${\CW} ({\vec l})$, for different ${\vec l}$'s (note that
as we change ${\vec l}$ we change the loop  in ${\CW} ({\vec l})$), we can
determine the positions of the vortices with arbitrary accuracy.

Alternatively we can  use the \nc analogue of the local energy density,
as constructed in \davakisun, to explore the multi-vortex configuration.
This operator  carries momentum $p$ and is given by
\eqn\energ{ \CE(p) = {\Tr}\left(\left(\exp\CD({\t}^{\m\n} p_{\n})\right) \,
F^2\right) \ , }
where a   straight Wilson line has been  inserted. In the
background of the multi-vortex solution,
$F^2= P_q$ and thus
$$\CE(p) = \sum_{i=0}^{q-1}{2\pi\over g^2 \t} e^{ip_{\n} x^{\n}_{i}} \ ,
$$
as if we have a vortex  of energy ${2\pi\over g^2 \t}$ localized at
the positions, $x_i^{\m} = {\t}^{\m\n}{\l}_{\n , i}$.
The Fourier transform of $\CE(p)$
would give an operator with local (delta-function) support at these positions,
however this is
not the Fourier transform of a given operator  since as $p$, and thus
$l$, changes
so does the length of the Wilson line.

There appears to be a contradiction between the behavior of the
current densities
that indicate that the vortices are spread out
in position space (over a size $~\sqrt{\t}$) and the behavior of the
Wilson loop or the gauge invariant energy density,
which indicate pointlike structure. We believe that the current densities
are more reliable, physical probes of the vortices.
This is because the current densities  have a cannonical normalization
(the $\Psi$'s are normalized eigenvectors), whereas the normalization
of  $\CE(p)$
or ${\CW} ({\vec l})$ is somewhat arbitrary. (In \davakisun
\ this problem was dealt with by considering ratios of
3 and 2-point functions of these operators). If we were to multiply
the loop operators by,
say $\exp[-p^2\t]$, this would not change the total energy or the
closed Wilson loop
and would produce the
expected form factors.

\subsec{Translating the vortices}

We should also be able to see that $\lambda_i$ correspond to the positions
of the vortices by performing a translation on the background
fields.
As we discussed before, the translations of  $D$ and $D^{\dagger}$
in the \nc plane
are generated by the gauge transformations and constant shifts of $A$.
  Thus, up to gauge tranformation, translations  by an amount $x_\mu$ simply
correspond to  shifts of
$D$ ($\bar D$)  by  $x/\t$ ($\bar x/\t $)--
which has the effect of shifting the $\lambda\mu$ by $ {\t}_{\n\m}x^{\m}$).
Thus,
as derived above,  the position of the i$^{\rm th}$ vortex is $x_i^{\m} =
{\t}^{\m\n}{\l}_{\n , i}$.

\subsec{Stability}

In the commutative two dimensional theory on an infinite plane one cannot
have stable localized droplets of magnetic flux. A vortex of quantized
flux, $Q$,
can be constructed by having the field equal to $Q/A$, in a region of area $A$.
The energy will be proportional to $(Q^2/A^2)A=Q^2/A$. This  simple
energy consideration
implies that a drop of flux will  immediately spread out  to fill all of
the space, and will have  vanishing field strength in  any finite region.
In the  \nc setup the  vortices that we have constructed
   are  classical solutions. But here too they are not stable--rather they
are metastable
and will decay and spread out if perturbed.

First of all, notice that the solutions with different magnetic charge can
be continuously
connected in field space. For example, let us take the charge one solution
$D=- S_1 c^{\dagger} {S_1}^{\dagger}$ and connect it to the vacuum solution
$- c^{\dagger}$ by a path:
\eqn\pth{
D_{\tau} = - \left( {\tau} S_1 c^{\dagger} {S_1}^{\dagger} +
(1- {\tau}) c^{\dagger} \right), \qquad {\tau} \in [ 0, 1]\ .
}
For every value of ${\tau}$ the gauge field
defined by \pth\ is well-defined,
\eqn\cong{ A_\tau= -\tau S_1[c^\dagger,  {S_1}^\dagger ] \ .}
Note that this would not  have  been the case if we decided, say,
to connect in the same vein the Dirac monopole and the trivial gauge field
on a two dimensional sphere.
Moreover, one can compute the flux of this gauge field, and its energy:
\eqn\flxeng{\eqalign{ {\Tr} F_{\tau} = {\tau}, \qquad  {\Tr} F_{\tau}^2& =
{\tau}^2 (2 - {\tau})^2 +
4{\tau}^2 ( 1- {\tau})^2 \sum_{m=1}^{\infty}
\left( {{2\sqrt{m}}\over{{\sqrt{m+1} + {\sqrt{m-1}}}}} - 1 \right)^2   \cr
& = {\tau}^2 \left[ ( 2 - {\tau})^2 + 4 a ( 1 - {\tau})^2  \right],
\qquad a \approx 0.173153 \ . \cr}   }
We see that the energy is finite for any $\tau$, that at $\tau = 0$
it has a minimum, and at ${\tau}=1$ it has a local maximum, while
it monotonically increases in between. We also see, that at ${\tau}=1$
there is a negative mode in the expansion of the energy around the
solution $D_{1}$.

Indeed, it can be verified, along the lines of \grossnekii\
that the spectrum of fluctuations around the $d$-vortex solution
contains a tachyonic mode. This mode was absent in the (essentially
the same) analysis of the fluctuations of the $d$-fluxons, presented
in \grossnekii\ due to the contribution of the Higgs field, which is
absent in the solution \gnrst. To see this expand about the vortex solution,
$$D=D_0+A, \quad \bar D=\bar D_0+\bar A; \quad \bar D_0=\lambda_q-S_q
c^\dagger{S_q}^\dagger \ .
$$
The quadratic part  of the action becomes
\eqn\quadr{ \CL_2 = \int dt  {\Tr} \left\{ -2 \partial_t A   \partial_t{\bar A}
-2P_q[\bar A, A] -([\bar D_0, A] -[D_0, \bar  A])^2
\right\} \ .
}
As in \grossnekii\ the fluctuating field $A$ can be decomposed into
four pieces, which would correspond to  modes due to $0-0$, $0-2$,$2-0$,$2-2$
strings respectively.  The $0-0$ modes, for example, correspond
to modes that lie in the $d$-dimensional subspace $V_q$, namely
$$A=\sum_{i,k=0}^{d-1} a_{ik}\vert i\rangle \langle k \vert , $$
These modes contribute to $ \CL_2$ the terms:
\eqn\zerzer{\int dt\sum_{i,k=0}^{d-1} \left\{ 2 \vert \partial_t
a_{ik} \vert^2
-\vert  a_{ik} \vert^2(\lambda_i-\lambda_j)^2  \right\}  \ ,
}
This is the quadratic piece
of the $0+1$ dimensional field theory of the $D_0$ branes, at
positions $\lambda_i$.
The diagonal components, $a_{ii}$, are the translation zero modes; whereas the
off-diagonal terms  acquire a (higgs) mass for separated vortices.

The unstable mode is actually in the off diagonal sector of the Hilbert space.
Consider the single vortex, $d=1$. In that case  it is easy to verify that
the mode: $ A= \vert1\rangle \langle0\vert b(t)$,
satisfies the equation of motion:    $\ddot b=b$, and is tachyonic.

A nice picture of the fluctuation spectrum can also be seen
by restricting to a subspace of ${\CF}$, which consists of the gauge fields
which have the form:
$$
D = - f(N) c^{\dagger}, \qquad N = c^{\dagger}c
$$
Such gauge fields form an invariant subspace with respect to the time
evolution  generated by the gauge theory Hamiltonian.
Indeed, the field strength is diagonal.
The potential energy becomes a functional on $f(N)$:
\eqn\ptn{V = \sum_{n=0}^{\infty} ( n \vert f(n) \vert^2
- (n+1) \vert f(n+1) \vert^2 +1 )^2 \ , }
Introduce $x_{n} = n \vert f(n) \vert^2$. The $d$-vortex solution
has:
\eqn\dvrtx {x_{n} = 0, \quad n \leq d; \qquad x_{n} = n  - d, \quad n > d\ , }
Expanding $V$ around this solution we get:
\eqn\expns{V = d - 2x_{d}^2 + O(x^4)\ , }
and we can identify $f(d)$ as the tachyon mode.
For $d=1$ this mode coincides with the mode described above.

\subsec{Unstable solutions and exact path integrals}

The $d$-vortex solutions are similar to the unstable monopole
solutions of  the Yang-Mills equations on a two dimensional
sphere, for the gauge group $SU(N)$. Consider, for simplicity, the
gauge group $SU(2)$. All the solutions are classified by a
non-negative integer $d$, and have the form: $$ A = d
\pmatrix{A_{Dir} & 0 \cr 0 & - A_{Dir} \cr} \ , $$ where $A_{Dir}$
is a constant curvature $U(1)$ gauge field on a two-sphere (which
can be obtained by restricting the Dirac monopole to the sphere).
The action of such a solution is: $$ S_{d} = {{2{\pi}^2
d^2}\over{g^2 A}} . $$ The partition function, as a function of
the area $A$ of the sphere and the coupling constant $g$, is given
by \tdgt: \eqn\tdimg{Z = \sum_{n=1}^{\infty} n^2 e^{-{\half} g^2 A
n^2} = {{\sqrt{2\pi}}\over{ g^3 A^{3\over 2}}} +
\sum_{d=1}^{\infty} {{\sqrt{8\pi}}\over{ g^3 A^{3\over 2}}} \left(
1 - {{4{\pi}^2 d^2}\over{g^2 A}} \right) e^{-S_{d}} } that is, for
this theory, the semi-classical approximation, together with a
finite number of quantum fluctuations, is exact, provided one sums
over all critical points of the action.

Let us now discuss the partition function of the two dimensional
Euclidean \nc gauge theory.
It is given by:
\eqn\prnct{Z = \int {\CD} D {\CD} D^{\dagger}
{\exp} -\left\{ {{{\pi}\t}\over{2g^2}}
   {\Tr} \left( [ D, {\bar D}] - 1\right)^2  \right\} \ .}
We immediately see, that if we were to apply the WKB approximation to this
theory, the partition function would diverge, for the unstable critical
points given by \gnrst\ have moduli, given by the eigenvalues of ${\bf \l}_i$,
and integrating along these moduli would render partition function divergent.
We propose to regularize this partition function by adding a gauge invariant
term
$$
{\ve} {\Tr} DD^{\dagger}\
$$
to the action. This is the analogue of the infrared regularization
provided by the
area of the two-sphere in the commutative example.

The related model \eqn\mtrx{Z_{\ve} = \sum_{N} e^{{\m} N}
\int_{Mat_{N \times N}}  {{{\CD} D {\CD} D^{\dagger}}\over{{\rm
Vol}(U(N))}} {\exp} - \left\{{{{\pi}\t}\over{2g^2}}
   {\Tr} \left( [ D, {\bar D}] - 1\right)^2 +{\ve} {\Tr} DD^{\dagger}
\right\} \ , } is extremely rich and can be solved exactly \kkn.
Whether this solution can lead to an exact solution of $2d$ NC
Yang-Mills theory
is a question that deserves further study.

\subsec{Supersymmetric solutions}
In the supersymmetric gauge theory, in addition to the gauge fields
we have fermions and scalars. The above analysis
is easily extended to include
these fields.
Let us discuss the case of the maximally supersymmetric theory.

In addition to the gauge fields, entering $D, {\bar D}$ we have a
collection of 7 (for the 2+1 dimensional theory) scalar fields
${\Phi}_{a}$, corresponding to the 7 transverse directions to D3
branes. These transform in the adjoint representation and thus are
to be  identified with   Hermitean operators ${\Phi}^a$ in the
Fock space ${\CH}$. The full bosonic part of the action is given
by (in the $A_t = 0$ gauge): \eqn\bsn{\eqalign{S =  {{2\pi
\t}\over{4 g^2}} \int {\rm d}t \, {\Tr}  & [ 4 {\dot D} {\dot{\bar
D}} + \sum_{a} {\dot {\Phi}^a} {\dot \Phi^{a}}  + \cr & \sum_{a}
[D, {\Phi}_{a}][{\bar D}, {\Phi}_a ] + \sum_{a\neq b} [ {\Phi}_a,
{\Phi}_b]^2 + 4 \left( [ D, {\bar D}] - {1\over{2\t}} \right)^2 ]\
. \cr}}

It is convenient to unify the Higgs fields and the gauge fields
into a single ${\bf 10}$-plet of (anti-Hermitean) operators
$D_{A}$, $A=0, \ldots 9$, acting in some Hilbert space ${\bf H}$.
The gauge theory is then described by the IKKT action \ikkt:
\eqn\mtrx{S = -{1\over{4g^2}} \sum_{A < B} {\Tr}_{\bf H} \left(
[D_A, D_B] + i {\t}_{AB} \right)^2 + \, fermions\ .} The equations
of motion following from \mtrx\ are (setting all fermions to
zero): \eqn\eommtrx{\sum_{A} [D_{A}, [ D_{A}, D_{B}]] = 0\ .} A
special class of solutions to \eommtrx\ are provided by the field
configurations that obey a stronger condition than \eommtrx: $$
[D_A, [D_B, D_C]] = 0\ , $$ for any $A, B,C$. Repeating the
arguments in \eom,\enrg,\pos,\ we arrive at the following generic
classification of the solutions with the finite $p$-tension:
\eqn\sltngen{[D_A, D_B] = i \left( - {\t}_{AB}  - f_{AB} P_{AB}
\right)\ ,} where $P_{AB} = P_{BA} = P_{AB}^{\dagger} $ are
projectors in ${\bf H}$, and $f_{AB}$ are  c-numbers . The
collection of the projectors $P_{AB}$ must have the following
properties: $$ [D_{C}, P_{AB}] = 0 \ , $$ $$ \sum_{A<B} f_{AB}^2
\, {\Tr}_{\bf H} P_{AB}^2 \sim V_{p}\ , $$ where $V_{p}$ is the
volume of the $p$-brane. The operators $D_A$ generate a certain
algebra, whose {\it spectrum} coincides with the worldvolume of
the configuration of the D-branes that the solution \sltngen\
corresponds to.

For example, the vacuum solution corresponding to a single flat
Dp-brane, extended in the directions $0,1, \ldots, p$, without any
$B$-field (i.e. ${\t}_{AB} = 0$) is: $$ D_{\m} = {\p}_{\m}, \,
{\m} = 0, \ldots, p; D_{A} = i x^A, \, A > p\ , $$ where $x^A$ are the
coordinates of the brane in the transverse space. The Hilbert
space is in this case the space ${\bf H} = S({\bf R}^{1,p})$ of
smooth functions of $\left( x^0, \ldots, x^{p}\right)$. If we were
to consider $N$ parallel flat branes then ${\bf H} = S({\bf
R}^{1,p} \times \{ 1, \ldots, N \})$, and the solution would
involve $9-p$ commuting $N \times N$ matrices ${\Phi}^A$, $A = p,
\ldots, 9$. The spectrum consists in this case of $N$ copies of
the space ${\bf R}^{1,p}$ (the latter emerges as the set of
eigenvalues of the operators ${\p}_{\m}$).

In general the solution \sltngen\ represents a collection of
D$p^{\prime}$-branes of various dimensionalities $p^\prime$. As
usual, one can read off the D-brane charges from the Chern
character: $$ {\Tr}_{\bf H}  {\exp} {1\over{2\pi i}} i f_{AB} dx^A
\wedge dx^B  P_{AB}\ . $$

The solution representing a D2-brane and a collection of $q$
D0-branes located at various points in the nine-dimensional space,
with $B$-field being in the $12$ directions,  is given by:
\eqn\twoplusone{\eqalign{& D_0 = {\p}_0 \ , \cr D_A = {\l}_A + i
{\t}_{AB} & S \, x^B \, S^{\dagger}, \qquad A = 1, \ldots, 9 \ ,\cr}}
with ${\l}_A$ being a diagonal matrix in the $N$ dimensional
subspace $V$ of the Hilbert space ${\bf H} = S({\bf R}^{1,0})
\otimes {\CH}$. The projectors $P_{AB}$ are given by: $$ P_{12} =
P_{21} = \int {\rm d}t \sum_{l=0}^{N-1} \vert l \rangle \langle l
\vert\ ,$$ with the rest vanishing.

Another interesting static solution, representing branes of
different dimensions, is given by: \eqn\threeplusone{ \eqalign{
D_0 = {\p}_0, \quad & \quad D_3 = {\p}_3 \cr D_A = {\l}_A + {\m}_A
x^3 + i {\t}_{AB} & S \, x^B \, S^{\dagger}, \qquad A = 1,2, 4
\ldots, 9 \cr}} with ${\l}_A, {\m}_A$ being the diagonal matrices
in the $N$ dimensional subspace $V$ of the Hilbert space ${\bf H}
= S({\bf R}^{1,1}) \otimes {\CH}$. In this case we see a D3-brane,
extended in the $0,1,2,3$ directions, with $1,2$ directions being
noncommutative, and a collection of $N$ D1-strings, forming various angles
with the D3-brane. This solution has $16N$ moduli (of which  $\sim
{\Tr} {\l}_1, {\Tr} {\l}_2$ can be eliminated by a gauge
transformation generating translations).

The stability of these solutions is analyzed by diagonalizing the
operator of quadratic fluctuations about the solution: \eqn\qudfl{
{\d}^2 S \, a_{A} = - \sum_{B} [D_{B}, [D_{B}, a_{A}]] + 2
[[D_{B}, D_{A}], a_{B}]\ , } where ${\d} D_{A} = a_{A}$ and the gauge
condition \eqn\ggcnd{[D_{A}, a_{A}] = 0\ , } has been imposed.

In
 the case of 0-2, 1-3 systems the only projectors involved were
 the projectors $P$ onto the $N$-dimensional subspace $V$ of the
 Hilbert space ${\bf H}$. Every operator  ${\CO}$ in ${\bf H}$ is
 canonically decomposed into four components:
\eqn\flucdec{\eqalign{{\CO} = & {\CO}^{VV} + {\CO}^{VH} +
{\CO}^{HV} + {\CO}^{HH} \cr & {\CO}^{VV} = P {\CO} P, \cr &
{\CO}^{VH}  = P {\CO} ( 1 - P), \, {\CO}^{HV} = (1- P) {\CO} P,
\cr & {\CO}^{HH} = (1- P){\CO}(1-P) \cr}} The fluctuation modes
${\CO}^{HH}$ are identical to the massless fields propagating on a
single D2 (D3) brane. The modes ${\CO}^{VV}$ coincide with the
fields of a matrix model describing $q$ D0-branes or with those of
rank $q$ matrix strings.

The modes ${\CO}^{VH}, {\CO}^{HV}$ are the interesting ones: they
contain tachyons (for unstable configurations) corresponding to
the decay of the lower-dimensional branes inside of the D2 (D3).
They also sometimes contain extra massless modes, responsible for
breaking of D1 strings into two semi-infinite strings, ending on
D3-brane \grossnekii.

\subsec{Plane wave solutions}

In the case of 1-3 system one expects tofind exact  plane waves
propagating along the string worldsheet.

Indeed, the following exace solution generalizes \threeplusone\ and
describes a collection of $q$ D1 strings with plane wave excitations
propagating along them: \eqn\kicks{\eqalign{& D_0 = {\p}_0, \qquad
D_3 = {\p}_3\ , \cr & D = f - S c^{\dagger} S^{\dagger}\ , \cr & D_{a} =
i f_a \ , \cr & f = \sum_{l=0}^{q-1} f_l (x^0, x^3) \vert l
\rangle\langle l \vert, \quad f_a = \sum_{l=0}^{q-1} f_{a,l} (x^0,
x^3) \vert l \rangle\langle l \vert \ , \cr & \left( {\p}^2_0 -
{\p}^2_3 \right) f = 0, \qquad a = 4, \ldots, 9 \cr & \left(
{\p}^2_0 - {\p}^2_3 \right) f_a = 0 \ .\cr}}

These solutions are not BPS and might be unstable to radiating 3-3
strings into the bulk.
The stability of these solutions will be analyzed elsewhere.

\subsec{Static strings}

The equations of motionthat follow from \bsn\ (setting all the
fermions to zero), for static fields are:
\eqn\eomb{\eqalign{& [D, [{\bar D}, {\Phi}_{a}]] + \sum_{b \neq a}
[ {\Phi}_{b}, [{\Phi}_{b}, {\Phi}_{a}]] = 0\ , \cr & [ {\Phi}_a ,
[ {\bar D}, {\Phi}_a]]  + [ {\bar D}, F] = 0 \ ,\cr & F = [ D,
{\bar D}] - 1 \ .\cr}} These equations are easily solved. Choose
$D$ and $\bar D$ to be as before, namely as in \gnrst. Then the
equations for the scalars reduce to: \eqn\eomb{\eqalign{&  \sum_{b
\neq a} [ {\Phi}_{b}, [{\Phi}_{b}, {\Phi}_{a}]] = 0\ , \cr & [
{\Phi}_a , [ {\bar D}, {\Phi}_a]]   =[ {\Phi}_a , [ { D},
{\Phi}_a]] = 0  \ .\cr}} By choosing $${\Phi}_a = {\underline
\l}_{q}^{a} = \sum_{i=0}^{q-1} {\l}_i^{a} \vert i \rangle \langle
i \vert ,$$ we clearly obey all of these equations. In addition we
can, of course shift all the $\Phi_a$ by multiples of the
identity.

These solutions can be lifted to higher dimensions, where more
parameters appear. For example, in the 3+1 dimensional theory, on
a \nc space with coordinates $t, x^3$ commuting and $x^1, x^2$
noncommuting, the operator ${\Phi}_7$ can be regarded as a gauge
field in the $3$  direction: $$ {\Phi}_7 = {\p}_3 + A_3\ ,$$ and
one can solve \eomb\ by setting \eqn\flxns{\eqalign{& A_3 = 0\ ,\cr &
{\Phi}_a = {\underline\l}_{q}^{a} + {\underline\m}_q^a \, x^3 ,
\qquad a = 1, \ldots, 6\cr & D = {\l}_q + {\m}_q \, x^3 \, - \,
S^{\dagger} c^{\dagger} S  \ , \cr}} where ${\underline\m}_q^a$
are a set of diagonal $q\times q$ matrices in $V_q$.
  This solution describes
a collection of $q$ D1-strings forming different angles with
D3-brane, set by the eigenvalues of the operators  ${\underline
\m}^a_q, {\m}_q$. In the following we assume, for simplicity, that
${\m}_q = 0$.

Let us define an operator $\vert {\m}_q \vert $ as the diagonal $q
\times q$ matrix:
\def\muq{\vert {\m}_{q} \vert}
$$
{\muq} = \sqrt{\sum_{a=1}^6 \left( {\m}_q^a \right)^2} \ .
$$
If all the D1-strings are parallel  to each other and form a
critical angle with the D3-brane then the solution \flxns\ describes
a general BPS $q$-fluxon of \grossnekii, with all moduli turned on
(this solution was announced in \grossnekii). In particular it
solves Bogomolny equation:
$$
B_{i} +  D_i {\Phi} = 0, \quad i=1,2,3; \qquad {\Phi}^a = {\l}^a_q +
n^a {\Phi} \ ,
$$
where $\sum_a ({n}^a)^2 = 1$, which means that
$ {\m}^a_q = 2 n^a \cdot {\rm I}_{q} $. This solution is stable.

However, for generic values
of ${\m}_d$ the solutions have negative modes.
  We shall now analyze these instabilities. As in \grossnekii\ the
fluctuations around the solution \flxns\
split into 1-1, 1-3, 3-1, and 3-3
sectors according to the decomposition of an arbitrary operator ${\CO}$
acting in $\CH$:
\eqn\dcmps{{\CO} = P_q {\CO} P_q + P_q {\CO} ( 1 - P_q) +
(1-P_q) {\CO} P_q + (1-P_q){\CO}(1-P_q)\ .}
Let us denote by $a_{\m}$ the fluctuations of $A_{\m}$ and by ${\vf}^a$
the fluctuations of ${\Phi}^a$, and by $X = {\half}(a_1 +i a_2)$ the
fluctuations of $D$:
\eqn\flgr{X = {\d} D, \quad a_3 = {\d} A_3 = - a_3^{\dagger}, \quad {\vf}^a =
{\d} {\Phi}^a = {\vf}^{a  \dagger}\ .}
We can split the fluctuations of the scalars into those that are transverse to
the strings and into   longitudinal fluctuations:
$$
{\vf}^a = {\bar\vf}^a +
( {{\m}^a}_q {\z} + {\z} {\m}^a_q ) \ ,
$$
where $\sum_a {\m}^a {\bar\vf}^a = \sum_a {\bar\vf}^a {\m}^a = 0$. The operator
${\z}$ can belong to the 1-1, 1-3, or the 3-1 sectors. Define  the operator
$Y$ to be:
$$
Y = a_3 + \muq\z + \z\muq \ .
$$

The 3-3 modes, corresponding to strings attached to the D3 brane,
  are identical to those in \grossnekii.
The instabilities appear in the
   1-3 and 3-1 sectors.
They have the following equations of motion
(cf. with Eq. (4.7) in \grossnekii. Note that
we  have  imposed  Lorentz  gauge
conditions on the fluctuations):
\eqn\eomfl{\eqalign{1-3: \qquad  & \left( {\p}_t^2 - {\p}_3^2 + 2(2{\hat n}+1)
+ \sum_{a} ({\l}^a_q + {\m}^{a}_q x_3)^2
+ 4 \right) X = 0 \ ,\cr
& \left( {\p}_t^2 - {\p}_3^2 + 2(2{\hat n}+1)
+ \sum_{a} ({\l}^a_q + {\m}^{a}_q x_3)^2  + 2 \muq
  \right) Y = 0\ , \cr
& \left( {\p}_t^2 - {\p}_3^2 + 2(2{\hat n}+1)
+ \sum_{a} ({\l}^a_q + {\m}^{a}_q x_3)^2
  \right) {\bar\vf}_b  = 0\ .\cr
3-1: \qquad &  \left( {\p}_t^2 - {\p}_3^2 \right) X + X \left( 2(2{\hat n}+1)
+ \sum_{a} ({\l}_q^a + {\m}_{q}^a x_3)^2
- 4 \right)  = 0 \ ,\cr
& \left( {\p}_t^2 - {\p}_3^2 \right) Y  + Y \left( 2(2{\hat n}+1)
+ \sum_{a} ({\l}_q^a + {\m}_{q}^a x_3)^2 - 2 \muq
  \right) Y = 0 \ ,\cr
& \left( {\p}_t^2 - {\p}_3^2  \right) {\bar\vf}_b + {\bar\vf}_b \left(
2(2{\hat n}+1)
+ \sum_{a} ({\l}_q^a+ {\m}_{q}^a x_3)^2
  \right)   = 0 \ . \cr}}
where
$$
{\hat n} \vert i \rangle \langle \psi \vert =  \vert i \rangle \langle \psi
\vert \left( - q +
(c^{\dagger} - {\bar\l}_i)(c - {\l}_i)\right), \quad i = 0, \ldots, q-1
$$
$$
{\hat n}^{\dagger} = {\hat n} \ .
$$

The resulting   spectrum of the 1-3, 3-1 fluctuations is:
\eqn\spctr{\eqalign{& {\p}_t^2 = - {\o}^2, \qquad
{\o}^2_i  = m_{i} + {\ve}_{i} \   \cr
m_{i} \, = \, & \left( \sum_{a}  ({\l}_i^{a})^2 \right)
- \left({1\over{{\muq}_i}} \sum_{a} {\l}_i^{a}{\m}_i^{a}\right)^2\  \cr
  1-3: \qquad   & X:\, {\ve}_i = {\muq}_i ( 2 m +1 ) + 2 ( 2n+3 ) \ \cr
& Y:\, {\ve}_i = {\muq}_i ( 2m+3) + 2(2n+1) \  \cr
& {\bar\vf}_a:\, {\ve}_i = {\muq}_i  ( 2 m +1 ) + 2 ( 2n + 1 ) \  \cr
3-1: \qquad & X:\, {\ve}_i = {\muq}_i ( 2 m +1 ) + 2 ( 2n-1)\   \cr
& Y:\, {\ve}_i = {\muq}_i ( 2m -1) + 2(2n+1)\  \cr
& {\bar\vf}_a:\, {\ve}_i = {\muq}_i  ( 2 m +1 ) + 2 ( 2n + 1 )\ , \cr}}
with $m,n \geq 0$.

The quantity $m_i$ determines whether the $i^{\rm th}$ th D1-string pierces
the D3-brane. In a 9 dimensional space a line and
a 3-plane do not intersect in general. The shortest distance between
them is $\sqrt{m_i} \geq 0$.
Thus we see that if for some $i=0, \ldots, q-1$ $\vert {\muq}_i -  2 \vert
> m_i$
then there is a tachyonic mode, either for $X$ (if ${\muq}_i < 2$), or
for $Y$ (if ${\muq}_i > 2$).

\newsec{U(2) monopoles}
\lref\adhm{ADHM} \lref\ho{P.~-M.~Ho, hep-th/0003012}
So far all of
our discussion has been devoted solely to the $U(1)$ \nc gauge theories that
arise in the Seiberg-Witten ${\ap}\to 0$ limit of   Dp-brane
theories with a $B$-field turned on.
The solitons we constructed were localized in the noncommutative directions,
but generically occupied all of the commutative space,
corresponding to   (semi)infinite D(p-2)-branes, immersed in  a Dp-brane,
or piercing it.
We now describe solitons which, although they have  finite
extent in the commutative directions, are nevertheless localized
and look like codimension three objects when viewed from far away. The
simplest such object is the monopole in the \nc $U(2)$ gauge
theory, i.e. the theory on a stack of two separated  D3-branes  in the
Seiberg-Witten limit \witsei.

We are interested in the $U(2)$ gauge theory on the \nc three
dimensional space. Let $H \approx {\bf C}^2$ be the Chan-Paton
space, i.e. the fundamental representation for the commutative
limit of the gauge group, and let $e_{0}, e_{1}$ denote an
orthonormal basis in $H$. The \nc version of the fundamental
representation is infinite dimensional, isomorphic to $H \otimes
{\CH}$. That is, the $U(2)$ matter fields ${\Psi}$ belong to the
space ${\CH} \otimes Fun (x^3) \otimes \left( {\CH} \otimes H
\right)$, where the first two factors make it a representation of
the algebra ${\CA}_{\t}$ of \nc functions on  ${\bf R}^3$, while
the second two factors make it a representation of the $U(2)$ \nc
gauge group. Actually, the latter is isomorphic to the group of
($x^3$-dependent) unitary operators in the Hilbert space ${\CH}
\otimes H$. Now, the Hilbert space ${\CH} \otimes H$ is isomorphic
to ${\CH}$ itself:
\eqn\iso{ \vert n \rangle \otimes \, e_{\a}
\leftrightarrow \vert 2n + {\a} \rangle\ .}
We wish to solve
Bogomolny equations:
\eqn\bgmlnii{[D_i, {\Phi}] = {i\over{2}}
{\ve}_{ijk} [D_j, D_k] - {\d}_{i3} {\t}\ , }
where ${\Phi}$ and $D_i$,
$i=1,2,3$ are the operators in ${\CH} \otimes H$ which have a
non-trivial magnetic charge:
\eqn\mgntch{Q_{m} = \int dx^3
{\Tr}_{\CH} {\p}_i \left( {\Tr}_{H} {\Phi} B_i \right)\ , } where $$
B_i = {i\over 2} {\ve}_{ijk} [D_j, D_k] - {\d}_{i3} {\t} \ ,$$ and
the Higgs field ${\Phi}$ approaches $$ \pmatrix{ a_+ & 0 \cr 0 &
a_{-}} \otimes {\rm I}_{\CH}$$ as  $x_3^2 + 2{\t} c^{\dagger} c
\to \infty$.

\subsec{Nahm's equations}

It was found in the study of commutative gauge theories that the
BPS solutions of gauge theory can be found via a sort of Fourier
transform, or reciprocity transformation \cg. In the instanton
case the four dimensional anti-self-duality equations are mapped
to matrix ADHM equations \cg. In the monopole case the three
dimensional equations go over to  one dimensional matrix
differential equations - Nahm's equations \nahm. As explained,
e.g. in \manuel, Nahm's equations are the BPS equations for
D1-strings suspended between D3-branes. (At the same time the ADHM
equations are analogous equations for D(-1)-branes dissolved
inside  D3-branes). In \hashimoto\grossneki\ the  \nc version of
Nahm's equations was derived. They have the form: \eqn\nahmm{
{\p}_{z} T_{i} = {i\over 2} {\ve}_{ijk} [ T_{j}, T_{k} ] -
{\t}{\d}_{i3}\ , } where in the case of the $U(2)$ gauge theory
the matrices $T_i$ have   size $k \times k$, with $k$ being the
monopole charge, the parameter $z$ takes values in the interval $I
= [a_{-}, a_{+}]$, and $T_l$ have   first order poles at the ends
of $I$ with residues forming a $k$-dimensional irreducible
representation of $U(2)$.

For $k=1$ this  means that the matrices must be regular
everywhere, which in turn yields  the unique solution:
\eqn\kone{T_{i}(z) = {\t} {\d}_{i3} z + {\kk}_{i}\ ,} where
${\kk}_{i}$ are arbitrary constants. This solution represents a
tilted D1-string suspended between two D3-branes, located at $z =
a_{-}$ and $z = a_{+}$ respectively.

The next step is to find a two-component spinor vector-function
$${\Psi} (z, {\vec x}) = \pmatrix{ {\Psi}_{+} \cr {\Psi}_{-} }\ ,
$$ which obeys the equation: \eqn\drc{ {\p}_z {\Psi} = i {\s}_{i}
\left( T_{i} (z) + x_i \right) {\Psi} \ ,} where ${\s}_i$ are the
Pauli matrices and $x_i$ are the spatial coordinates,  i.e. the
generators of ${\CA}_{\t}$. The fundamental solution to \drc\ is a
$k \times 2$ spinor-valued matrix (both ${\Psi}_{+}$ and
${\Psi}_{-}$ are $k\times 2$ matrices whose entries belong to
${\CA}_{\t}$). The solution to \drc\ is defined up to   right
multiplication by an element of ${\rm Mat}_{2}({\CA}_{\t}) \approx
{\CA}_{\t} \otimes {\rm End}(H)$. Among  these elements the
unitary elements (i.e. the ones which solve the equation
$uu^{\dagger} = u^{\dagger}u = 1$) are considered to be the gauge
transformations. In the commutative setup one normalizes ${\Psi}$
as follows: \eqn\nrmcm{\int \, dz \, {\Psi}^{\dagger} {\Psi} =
{\rm I}_{2} = \pmatrix{ 1 & 0 \cr 0 & 1}\ .}

Finally, given ${\Psi}$ the solution for the gauge and Higgs
fields is given explicitly by:
\eqn\gfhf{\eqalign{{\Phi} \quad = &
\quad \int \, dz \, z \, {\Psi}^{\dagger} {\Psi} \ ,\cr A_{i} \quad =
& \quad \int \, dz \, {\Psi}^{\dagger} {\p}_i {\Psi} \ .  \cr}}

For $k=1$, by shifting $x_i$ we can always set  ${\kk}_i =
0$.

\subsec{Commutative case}

We start with the commutative case, as it will be one of the
limits to which our solution reduces as ${\t} \to 0$. In the case
$k=1$ the analysis simplifies: $T_i = 0$, one can take $a_{\pm} =
{\pm}{a\over 2}$ and \eqn\fndm{{\Psi} = \pmatrix{ \left( {\p}_z +
x_3 \right) v \cr \left( x_1 + i x_2 \right) v }, \quad {\p}_z^2 v
= r^2 v, \quad r^2 = \sum_i x_i^2 \ .} The condition that $\Psi$
is finite at both ends of the interval allows for  two solutions
of \fndm: $$ v = e^{\pm r z}, $$ which after imposing the
normalization condition,\nrmcm, leads to: $$ {\Psi} = {1\over
\sqrt{2{\rm sinh} (ra)}} \pmatrix{ \sqrt{r + x_3} e^{rz} & -
\sqrt{r - x_3} e^{-rz} \cr {x_{+} \over \sqrt{r + x_3}} e^{rz} &
{x_{-} \over \sqrt{r - x_3}} e^{-rz} }, $$ where $x_{\pm} = x_1
{\pm} i x_2$.

In particular,
$$ {\Phi} = {1\over 2}\left( {a\over  {\rm tanh}
(ra)} - {1\over r} \right) {\s}_{3}. $$

\subsec{Noncommutative case}

In the case $k=1$ we take: $T_{1,2} =0, \quad T_{3} = {\t}z$.
Following \grossneki\ we introduce the operators: \eqn\bop{b =
{1\over{\sqrt{2\t}}} \left( {\p}_{z} + x_3 + {\t} z \right), \quad
b^{\dagger} = {1\over{\sqrt{2\t}}} \left( - {\p}_{z} + x_3 + {\t}
z \right)\ ,} which obey $[b, b^{\dagger}] = 1$. We also introduce
the superpotential $W$: \eqn\super{W = {1\over{2\t}}\left( x_3 +
{\t} z \right)^{2}\ ,} whose importance arises from the formulae:
\eqn\superb{b = {1\over{\sqrt{2\t}}} e^{-W} {\p}_z e^{W}, \quad
b^{\dagger} = - {1\over{\sqrt{2\t}}} e^{W} {\p}_{z} e^{-W}\ .} It
is convenient to choose units, where $2{\t} =1$.

Equations \drc\  then take the form: \eqn\drcnc{\eqalign{ &
b^{\dagger} {\Psi}_{+} + c {\Psi}_{-} = 0\ , \cr  -& c^{\dagger}
{\Psi}_{+} + b {\Psi}_{-} = 0 \ ,\cr}} where ${\Psi}_{\pm}(z) \in
{\CA}_{\t}$. It is convenient to solve first the equation
\eqn\drcnch{\eqalign{ & b^{\dagger} {\e}_{+} + c {\e}_{-} = 0\ ,
\cr  -& c^{\dagger} {\e}_{+} + b {\e}_{-} = 0 \ ,\cr}} with
${\e}_{\pm} (z) \in {\CH}$. The latter has the following
solutions: \eqn\slh{\eqalign{ {\ve}^{\a} \quad = & \quad
\pmatrix{{\e}_{+}^{\a} \cr {\e}_{-}^{\a} \cr}\ , \, \a = 0, 1 \cr
{\ve}_{0}^{0} \quad = \quad \pmatrix{ 0 \cr
{1\over{\sqrt{{\z}_0}}} e^{-W} \vert 0 \rangle \cr} \ , & \qquad
{\z}_0 = \int_{a_{-}}^{a_{+}} \, dz \, e^{-2W}, \, {\ve}_0^{1} = 0
\cr {\ve}_{n}^{\a} \quad =  \quad \pmatrix{ b \, {\b}_{n}^{\a}
\vert n-1 \rangle \cr \sqrt{n} \, {\b}_{n}^{\a} \vert n \rangle
\cr},\, & \qquad n > 0 \ ,\cr}} The functions ${\b}_n^{0},
{\b}_{n}^{1}$ solve \eqn\schroed{ \left( b^{\dagger}\, b + n
\right) {\b}_n^{\a} = 0,} and are required to obey the following
boundary conditions: \eqn\bndry{\eqalign{b{\b}_n^{1} (a_{+}) = 0,
\quad & \quad {\b}_{n}^{0} (a_{-}) = 0 \cr {\b}_{n}^{1} b
{\b}_{n}^{1} (a_{-}) = -1, \quad & \quad {\b}_{n}^0 b {\b}_{n}^0
(a_{+}) = 1 \cr}} A solution to \drcnc\ is given by:
\eqn\asln{\Psi = \sum_{n \geq 0, \, {\a} = 0,1 } {\ve}_{n}^{\a}
\cdot \langle n- {\a} \vert\otimes e_{\a}^{\dagger}  .} The
conditions \bndry\ together with  \schroed\ imply that: $$ \int\,
dz\, \left( {\ve}_{n}^{\a} \right)^{\dagger} {\ve}_{m}^{\g} =
{\d}^{\a\g} {\d}_{mn}, $$which in turn yield \nrmcm. All other
solutions to \drcnc\ are gauge equivalent to \asln.

It is easy to generate the solutions to \schroed: first of all, $$
f_{n}(z) = b^{n-1}(e^{W}) = e^{W(z)} h_{n-1}(2x_3 +z), \qquad
h_{k} (u) = e^{-{u^2 \over 4}} {d^{k}\over{{d u^k}}}  e^{u^2 \over
4}$$is a solution . Then $$ {\hat f}_n = f_n (z)\int^{z}
{{du}\over{f_{n}(u)^2}}$$ is the second solution. Notice that, for $k$ even,
$h_{k} (u) > 0 $ for all $u$   and, for $k$ odd, the only zero of
$h_{k}(u)$ is at $u=0$, and $h_{k}(u)/u > 0$ for all $u$.
Therefore, ${\hat h}_k (z)$ is well-defined for all $z$.

Consequently, \eqn\betas{\eqalign{& {\b}_{n}^{0}(z) = {\tilde\n}_n \,
f_n (z) \int_{a_{-}}^{z} {{du}\over{f_{n}(u)^2}} \ ,\cr &
{\b}_{n}^{1} = {\n}_{n}\,  \left( - {1\over{n f_{n+1}(z)}} + f_{n}
(z) \int^{a_{+}}_{z} {{du}\over{f_{n+1}(u)^2}} \right)\ , \cr} } where
\eqn\nus{\eqalign{{\n}_{n}^{-2} \quad = & \quad \left(
f_{n}(a_{-})f_{n+1}(a_{-}) \int_{a_{-}}^{a_{+}}
{{du}\over{f_{n+1}^2(u)}} - {1\over{n}} \right)
\int_{a_{-}}^{a_{+}} {{du}\over{f_{n+1}^{2}(u)}}\ , \cr
{\tilde\n}_{n}^{-2} \quad = &\quad \left(
f_{n}(a_{+})f_{n+1}(a_{+}) \int_{a_{-}}^{a_{+}}
{{du}\over{f_{n}^2(u)}} + 1 \right) \int_{a_{-}}^{a_{+}}
{{du}\over{f_{n}^{2}(u)}} \ .\cr }} (again, note that ${\b}^{\a}_n
(z)$ are regular at $z = - 2x_3$).

  We now are in position to calculate the components of
the Higgs field and of the gauge field. We start with
\eqn\hggs{\eqalign{{\Phi} = \int \, dz \, z \, {\Psi}^{\dagger}
{\Psi} = & \quad \sum_{n \geq 0, \, {\a},{\g} = 0,1}
{\vf}^{\a\g}_{n} \cdot e_{\a} e_{\g}^{\dagger} \otimes \vert n -
{\a} \rangle \langle n - {\g} \vert \, , \cr {\rm where} \quad
{\vf}^{\a\g}_{n} = & \int \, dz \, z {\ve}_{n}^{\a, \dagger}
{\ve}_{n}^{\g} = - 2x_3 {\d}^{\a\g} + \int (b{\b}_n^{\a}) (b +
b^{\dagger}) (b{\b}_n^{\g}) + n {\b}_n^{\a} (b + b^{\dagger})
{\b}_n^{\g}  \cr  \qquad\qquad = - 2x_3{\d}^{\a\g} & + \left(
(b{\b}_n^{\a})(b{\b}_n^{\g}) - n
{\b}_n^{\a}{\b}_n^{\b}\right)\vert_{a_{-}}^{a_{+}} \ .\cr}} The
component $A_3$ of the gauge field vanishes, just as in  the case of the
U(1) solution of \grossneki: \eqn\athree{A_3 = \int
{\Psi}^{\dagger} {\p}_3 {\Psi} = \int \left( (b {\b}_n^{\a})
{\p}_3 ( b {\b}_n^{\g} ) + n {\b}^{\a}_n {\p}_3 {\b}_n^{\g}
\right) \cdot e_{\a} e_{\g}^{\dagger} \otimes \vert n - {\a}
\rangle \langle n - {\g} \vert = {\half} {\p}_3 \int
{\Psi}^{\dagger} {\Psi} = 0 \ .}

The components $A_1, A_2$ can be  read off  from the expression for the
operator $D$: \eqn\dopr{\eqalign{D = -\int & \, dz \,
{\Psi}^{\dagger} c^{\dagger} {\Psi} \cr = & \sum_{n \geq 0,
{\a},{\g} = 0,1} D_{n}^{\a\g} \cdot \, e_{\a} e_{\g}^{\dagger}
\otimes \vert n+1 - {\a} \rangle \langle n - {\g} \vert\ , \cr {\rm
where} \, \, D_{n}^{\a\g} = & - \sqrt{n} \left( {\b}_{n+1}^{\a}
(b{\b}_{n}^{\g}) \right) \vert_{a_{-}}^{a_{+}}\ . \cr}}

The solution \hggs\dopr\ has several interesting length scales
involved (recall that our units above are such that $2{\t} =1$):
$$ {\t} \vert a_{+} - a_{-} \vert, \quad \sqrt{\t}, \quad
{1\over{\vert a_{+} - a_{-} \vert}\ .} $$

By shifting $x_3$ we can always assume that $a_{-} = 0, a_{+} = a
> 0$.

\subsec{Suspended D-string}

In this section we set $\t$  back to $\half$. As we discussed
before the spectrum of the operators $D_A$, $A = 0, \ldots, 9$
determines the ``shape'' of the collection of D-branes the
solution of the generalized IKKT model \cds\ corresponds to. To
``see'' the spatial structure of our solution let us concentrate
on the $\langle 0 \vert {\Phi} \vert 0 \rangle$ piece of the Higgs
field, for it describes the profile of the D-branes at the core of
the soliton. From \hggs\ we see that $$\langle 0 \vert {\Phi}
\vert 0 \rangle = \pmatrix{{\rho}_{+} & 0 \cr 0 & {\rho}_{-} \cr}
\ , $$ where ${\rho}_{+} = {\vf}^{00}_{0}, {\rho}_{-} =
{\vf}^{11}_{1}$.

Let us look specifically at the component ${\rho}_{+}$ of the
Higgs field: \eqn\snakeyes{\eqalign{{\rho}_{+} \, &= \, - {\half}
{{\p}\over{{\p} x_3}} {\rm log} \left( \int_{0}^{a} dp \, e^{-2x_3
p - {\half} p^2} \right)\ ,\cr & = - 2x_3 + \langle\langle p
\rangle\rangle_{2x_3}^{a + 2x_3}\ , \cr & = \, - 2x_3 - 2
{{e^{-{{(a + 2x_3)^2}\over 4}} - e^{-{{(2x_3)^2}\over
4}}}\over{{\g}( {\scriptstyle a + 2x_3}) - {\g}( {\scriptstyle
2x_3})}} \ ,\cr }} where \eqn\defs{ \langle\langle {\CO}
\rangle\rangle_{\a}^{\b} = {{\int_{\a}^{\b} {\CO} e^{-{p^2 \over
4}} \, dp}\over{\int_{\a}^{\b} e^{-{p^2 \over 4}} \, dp}} ,  \quad
{\g}(z) = \int_{0}^{z\over 2} dp \, e^{-{p^2 \over 4}}\ .  }
   The
$\langle\langle\ldots\rangle\rangle$ representation of the answer
helps to analyze the qualitative behavior of the profile of
$\rho_{+}$. Clearly, the truncated Gaussian distribution which
enters the expectation values $\langle\langle\ldots\rangle\rangle$
in \snakeyes\ favors $p \approx 0$ if ${\a} < 0 < {\b}$, $p
\approx {\a}$ for ${\a} > 0$ and $p \approx {\b}$ for ${\b} < 0$.
Thus, \eqn\spke{\eqalign{{\rho}_{+} \sim 0, & \qquad x_3 > 0 \cr
{\rho}_{+}\sim - 2x_3, & \qquad
   0 > x_3 > {-\half} a \cr
  {\rho}_{+} \sim a, & \qquad {-\half} a > x_3 \ . \cr}}

   This behavior agrees with the expectations about the tilted
   D1-string suspended between two D3-branes separated by a distance
   $\vert a \vert$. The eigenvalue ${\rho}_{+}$ corresponds roughly to the
    the transverse
coordinate of the D1 string, that runs from $a$ at large negative
$x_3$ to $0$ at large positive $x_3$.  In between the linear
behavior of the Higgs field corresponds to the D1 string tilted at
the critical angle. Indeed, for large $a \gg 1$,  in the region $0
> x_3
> {-\half} a$ this solution looks very similar to that of a single
fluxon \grossnekii.

For future reference let us present the expression for another
eigenvalue of $\langle 0 \vert {\Phi} \vert 0 \rangle$,
${\rho}_{-}$: \eqn\romin{\eqalign{{\rho}_{-} = & \quad {{2x_3
(2x_3 + a) M + M^2 - (2x_3 + a)^2 - e^{-2x_3 a -
{{a^2}\over{2}}}}\over{M ( 2x_3 + a - 2x_3 M)}} \cr \quad {\rm
where} & \quad M = e^{-2x_3 a - {{a^2}\over{2}}} + (2x_3 + a)
\int_{0}^{a} e^{-2x_3 p - {{p^2}\over{2}}} \, dp \cr}}

At this point, however, we should warn the reader that only the
eigenvalues of the full, $2 \infty \times 2 \infty$ operator
${\Phi}$ should be identified with the D-brane profile. The
components ${\rho}_{\pm}$ do not actually coincide with any of
them. The eigenvalues of ${\Phi}$, as it follows from the
representation \gfhf, are located between $0$ and $a$, which is
also what we expect from the dual D-brane picture \hashimoto.

\newsec{Conclusions}

    In this paper we have presented a rather complete description of
the classical, soliton, solutions of co-dimension two  in \nc gauge theory.
We showed that the \nc gauge theory contains the classical and quantum dynamics
of all $U(N)$ gauge theories and that classical solutions are labeled by
the rank of the gauge group and the magnetic charge.
We presented many examples of BPS and non-BPS solutions that can be
constructed from the basic
set of solutions when other matter fields are turned on. The BPS
solutions describe various D-1
strings attached or piercing D3 branes. We analzed how the non-BPS
solutions are unstable.

In addition we gave an explicit construction of a (localized in 3
dimensions) $U(2)$ monopole, which has an intricate and
interesting structure that corresponds precisely to the picture of
a monopole as being a finite D1 string attached to two separated
D3 branes.

  The various solitons we have analyzed should have an interesing S-dual
description in terms of fundamental strings, presumable in \nc open
string theory. For example,
if we wrap the  non-BPS fluxon (with constant $\Phi$) around a circle
in the commutative ($x_3$)
direction it should correspond in the strong coupling limit to a fundamental
  closed string wound around the circle.
The instability of the fluxon   to spread out over all  the \nc space,
should be the S-dual of the transition of the closed string to an
open string on the brane,
which can then dissipate.

Finally, we set up the machinery to   derive an exact analytic soltution of
2 dimensional \nc gauge theory. It would be of great interest to complete this
construction.

\bigskip
\bigskip
\bigskip
\ndt {\bf Acknowledgements.}

\ndt{}We would like to thank  A. Hashimoto and  N. Itzhaki for
discussions. Our research was partially supported by NSF under the
grants PHY 99-07949 and PHY 97-22022; in addition, research of NN
was supported by Robert H.~Dicke fellowship from Princeton
University, partly by RFFI under grant 00-02-16530, partly by the
grant 00-15-96557 for scientific schools. NN is also grateful to
ITP, UC Santa Barbara for hospitality while this project was
carried out.

\bigskip\bigskip\bigskip

\footatend\vfill\supereject\immediate\closeout\rfile\writestoppt
\baselineskip=14pt\centerline{{\bf References}}\bigskip{\frenchspacing%
\parindent=20pt\escapechar=` \input refs.tmp\vfill\eject}\nonfrenchspacing \bye